\documentclass[%
reprint,
superscriptaddress,
amsmath,amssymb,
aps,
prb
floatfix,
showpacs,
]{revtex4-2}

\usepackage{graphicx}
\usepackage{dcolumn}
\usepackage{bm}
\usepackage{hyperref}
\usepackage{ulem}

\usepackage[english]{babel}
\usepackage[utf8]{inputenc}
\usepackage{amssymb,amsfonts,amsmath,mathtools,mathrsfs}
\usepackage{relsize}
\usepackage{mdframed}
\usepackage[margin=0.6in]{geometry}
\usepackage{enumitem}
\usepackage{graphicx}
\usepackage{float}
\usepackage{url,hyperref}
\usepackage[table,usenames,dvipsnames]{xcolor}
\usepackage{gensymb}
\usepackage{grffile}

\hypersetup{
	colorlinks=true,
	linkcolor=Blue,
	filecolor=magenta,      
	urlcolor=Blue,
	citecolor=Blue,
}
\usepackage[bottom]{footmisc}
\usepackage[capitalize]{cleveref}
\usepackage{textcomp}
\usepackage{braket} 

\setcitestyle{line}

\usepackage{amsmath,amssymb,wasysym}

\makeatletter
\def\@fnsymbol#1{\ensuremath{\ifcase#1\or *\or \dagger\or \ddagger\or
   \mathsection\or \mathparagraph\or \|\or **\or \dagger\dagger
   \or \ddagger\ddagger \else\@ctrerr\fi}}
    \makeatother

\begin{document}

\title{Investigation of magnetic and transport properties of GdSbSe}

\address{Department of Chemistry, Indian Institute of Technology Delhi, New Delhi 110016, India}
\author{Aarti Gautam}
\address{Department of Chemistry, Indian Institute of Technology Delhi, New Delhi 110016, India}
\author{Prabuddha Kant Mishra}
\address{Department of Chemistry, Indian Institute of Technology Delhi, New Delhi 110016, India}
\author{Souvik Banerjee}
\address{Chemistry and Physics of Materials Unit, School of Advanced Materials,
Jawaharlal Nehru Centre for Advanced Scientific Research, Bengaluru 560064, India}
\author{A. Sundaresan}
\address{Chemistry and Physics of Materials Unit, School of Advanced Materials,
Jawaharlal Nehru Centre for Advanced Scientific Research, Bengaluru 560064, India}
\author{Ashok Kumar Ganguli}\email[E-mail: ]{ashok@chemistry.iitd.ac.in}
\address{Department of Chemistry, Indian Institute of Technology Delhi, New Delhi 110016, India}
\address{Department of Chemical Sciences, IISER Berhampur, Odisha-760003, India}

\begin{abstract}

We report the detailed investigation of the magnetic, transport, and magnetocaloric effects of   GdSbSe by magnetic susceptibility  $\chi(T)$, isothermal magnetization $M(H)$, resistivity $\rho(T, H)$, and heat capacity $C_p(T)$ measurements, crystallizing in the ZrSiS-type tetragonal crystal system with space group $P4/nmm$. Temperature-dependent magnetic susceptibility measurements revealed long-range antiferromagnetic ordering with two additional magnetic anomalies below Néel temperature ($T_N$ $\approx$ 8.6 K), corroborated through magnetocaloric and specific heat studies. Isothermal magnetization measurements unveil hidden metamagnetic signatures through a clear deviation from linearity. In addition, the enhanced value of the Sommerfeld coefficient ($\gamma$ = 152(5) mJ/ mol K$^2$) suggests strong electronic correlations in GdSbSe. The entropy of magnetization derived from magnetic isotherms unfolds the field-induced transition from Inverse magnetocaloric Effect (IMCE) to Conventional MCE. The detailed transport properties indicate a semimetallic behavior, strongly coupled with magnetic order. Deviations from Kohler's rule and non-linear Hall resistivity anticipate the possibility of Dirac-like dispersion with non-trivial characteristics.

\end{abstract}

\maketitle

\section{Introduction}

The discovery of topological insulators, with surface states protected by time-reversal symmetry(TRS) or specific crystalline symmetries\cite{Hasan2010, Fu2011}, has sparked the pursuit of topological materials, especially in semimetals\cite{burkov2016, Lv2021}. Topological semimetals(TSMs), characterized by their nature of band crossing near the Fermi levels have garnered special attention due to their exotic and strongly correlated properties\cite{Huang2015}. Unlike topological insulators, the band crossing in TSMs occurs in bulk rather than the surface, allowing the linear dispersion of bands in 3D space\cite{Armitage2018}.In Dirac and Weyl semimetals, the band crossing occurs at a single point\cite{Young2012} whereas, it extends along a continuous line in nodal line and a ring in nodal ring semimetals \cite{Burkov2011}. As established earlier, the topological states in nodal line semimetals without spin-orbit coupling (SOC) are protected through TRS and inversion symmetry. However, the system comprising of spin-orbit coupling with inversion symmetry alone cannot protect nodal line surface states, it requires additional crystal symmetries such as mirror plane symmetries or glide plane/axis to protect the surface states\cite{Fang2015}. Out of all the symmetries, non-symmorphic symmetry stands out due to its robustness against SOC, which provides an additional edge to explore topologically protected states in compounds with significant SOC\cite{Klemenz2019, Klemenz2020}.
\begin{figure*}
\begin{center}
\includegraphics[width= 2.05\columnwidth,angle=0,clip=true]{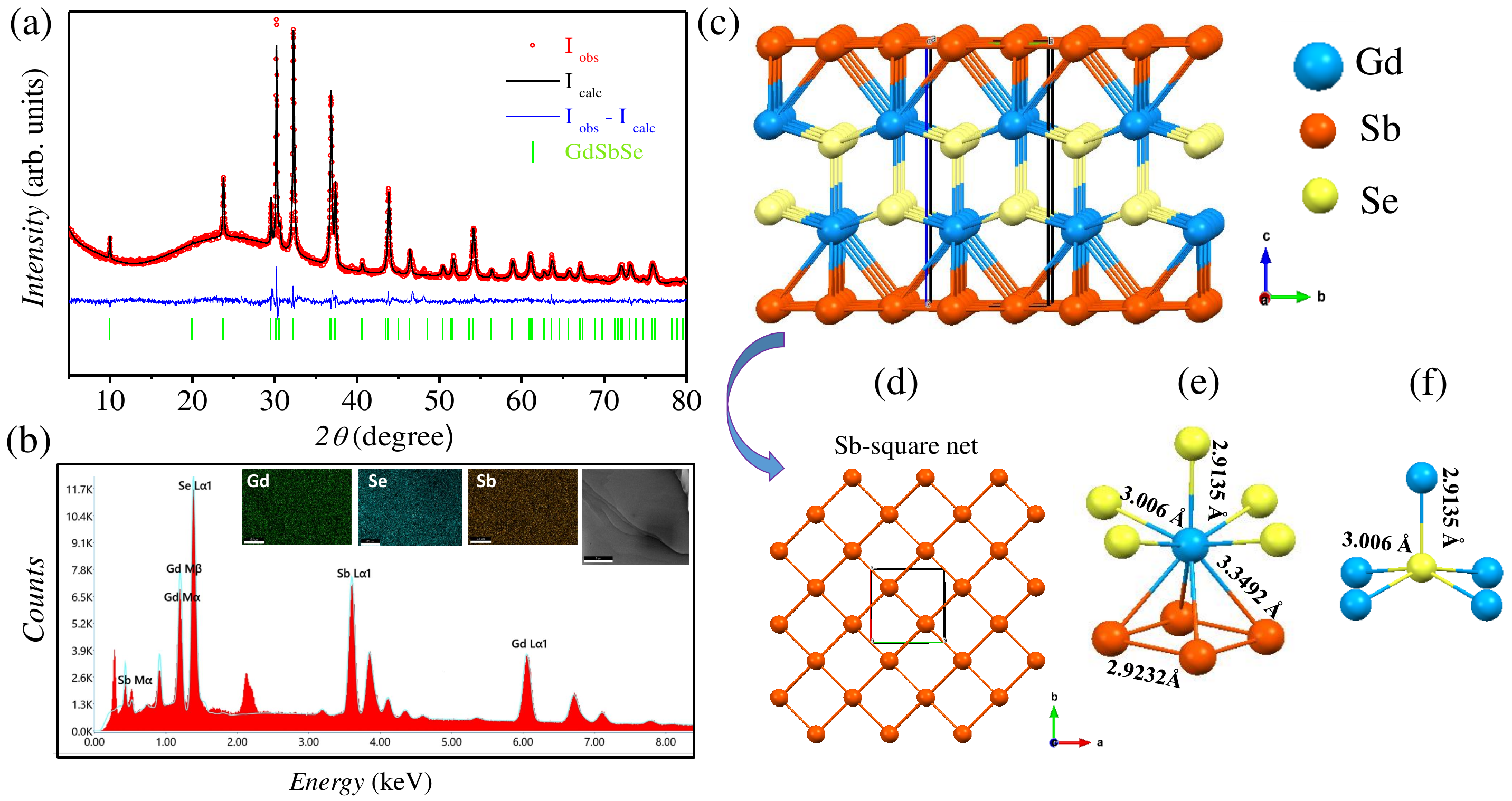}
\caption{(Color online) (a) Rietveld refined powder X-ray pattern of polycrystal of GdSbSe. The blue line represents the difference between the calculated and observed intensities. The green vertical bars indicate the allowed Bragg's reflections for GdSbSe.(b) EDX spectroscopy data for GdSbSe. The FESEM image in the inset shows the layered nature of the sample along with elemental mapping of constituent elements. (c) Crystal structure of GdSbSe displaying GdSe layers stacked with Sb layers along the c-direction. (d) square-net structure of Sb-layer viewed from c-direction. (e)  Nine-fold Gd-coordinated with four Sb and five selenium atoms. (f) Five-fold coordinated selenium with Gd atoms.}
\label{Fig. 1}
\end{center}
\end{figure*}
Within this captivating domain of topological nodal line semimetals, the WHM compounds (W = Zr, Hf, or La, H = Si, Sb, Ge, or Sb, and M =  S, O, Se, or Te)\cite{wang1995, Xu2015} have taken the front seat due to various interesting properties intertwined with their crystal symmetry. ZrSiS, one of the extensively studied members of the WHM family, hosting multiple Dirac cones with nodal-line semimetallic phase exhibits exotic properties such as giant magnetoresistance\cite{lv2016, Yang2021} and quantum oscillations\cite{Singha2017, Hu2017}. Recently, ARPES studies have established the linear dispersion of bands near the Fermi level with significant SOC in LaSbTe\cite{Wang2021}, whereas CeSbTe exhibits symmetric Dirac cones and magnetically tunable Weyl states\cite{lv2019}. Moreover, other rare-earth analogs with appreciable f-electrons host nodal-line states as demonstrated through ARPES in GdSbTe \cite{Hosen2018} and SmSbTe \cite{Regmi2022}. The LnSbX  (Ln = La - Lu, X = S, Se and Te) series feature complex magnetic states like the devil's staircase and metamagnetic transitions accompanied by Kondo scattering\cite{Chen2017, Pandey2020, lv2019}. Furthermore, signatures of charge density waves have also been probed in the non-stoichiometric composition (LnSb$_x$Te$_{2-x}$) of the LnSbTe (Ln = La, Ce, Gd, and Sm)\cite{Lei2019, Lei2021} series. Given, the rich magnetic and exotic topological properties of the LnSbX series of compounds, it provides an interesting platform to investigate the interplay of magnetism, correlation, and topology. Inspired by the effect of SOC on the tunability of nodal-line semimetallic phase in ZrSiS through chalcogen substitution\cite{Hosen2017, Song2021}, we have attempted to synthesize the selenium analog of GdSbTe which is established as a nodal line antiferromagnetic semimetal\cite{Sankar2019}. Interestingly, the selenium analog for CeSbTe and LaSbTe has been realized experimentally\cite{Pandey2022} but the properties for GdSbSe are not explored yet. Herein, we report the experimental investigation on the new family member of the non-symmorphic family crystallizing in ZrSiS type structure. We present the synthesis, magnetization, transport, and magneto-transport properties of GdSbSe.  Our experimental investigation on GdSbSe polycrystals shows the antiferromagnetic (AFM) ground states below T$_N$ $\approx$ 8.6 K. Field-dependent magnetization studies indicate the metamagnetic signatures below the transition temperature.
Furthermore, our transport and magneto-transport studies established a correlation between the magnetic and transport properties. Our observations from Hall measurements and magnetoresistance measurements unveiled non-trivial characteristics. Specific heat measurements provided insights into electron correlation enhancement with selenium substitution. Additionally, we conducted a deeper investigation into the mobility and nature of charge carriers. The aforementioned studies on this class of materials can pave new opportunities to study the impact of X (chalcogen-site) substitution on their physical properties.


\section{Experimental Details}

Polycrystalline GdSbSe was prepared via a conventional solid-state sealed tube reaction route. The elements were mixed well in their stoichiometric ratio (Gd:Sb:Se: 1:1:1), sealed in an evacuated quartz tube (10$^{-5}$ bar), and heated to 1123 K for 48h. The reaction mixture was again ground, pelletized, and sintered at 1123 K for 48h to ensure better phase homogeneity and ambient atmosphere stability. The phase purity was examined through powder X-ray diffraction pattern using Bruker D8 Advance Diffractometer with Cu-K$\alpha$ radiation. The structural refinement using the Rietveld method was performed with the TOPAS software package\cite{topas}. The compositional analysis and elemental mapping were carried out using the field emission scanning electron microscope (FESEM TESCAN, MAGNA). Temperature and field-dependent magnetization measurements were carried out using the Magnetic Property Measurement system) MPMS3 SQUID Magnetometer (Quantum Design) equipped with a 7 T magnet. The resistivity measurements as a function of temperature and magnetic field were performed using the standard four-probe technique at the Physical Property Measurement System (PPMS, Quantum Design). In addition, magneto-resistance and Hall resistivity have been symmetrized with respect to H = 0 T to remove the contribution due to misalignments of contacts. The specific heat measurements were also performed within the 2–100 K temperature range using a Physical Property Measurement System (PPMS, Quantum Design).





\section{RESULTS AND DISCUSSION}

\subsection{Crystal structure}
The Rietveld structural refinement as shown in \hyperref[Fig. 1]{Fig. 1}(a), was performed for the powder X-ray diffraction pattern observed at room temperature which reveals that GdSbSe crystallizes in the tetragonal crystal system with space group  $P4/nmm$. The extracted lattice parameters are a = b = 4.1340(2)(\r{A}) and c = 8.8870(6)(\r{A}), comparatively smaller than the previously reported tellurium analog of this compound and consistent with the smaller selenium atomic radii\cite{Sankar2019}. The Wyckoff positions with other refined parameters of the constituting elements are tabulated in Table \ref{table1}. The structure of GdSbSe can be described as the stacking of [Gd-Se] layers with [Sb] square-net layers along the c-direction, as illustrated in \hyperref[Fig. 1]{Fig. 1}(c) and \hyperref[Fig. 1]{Fig. 1}(d),respectively. \hyperref[Fig. 1]{Fig. 1}(e) shows the nine-fold coordination of Gd, which forms four equidistant bonds [Gd-Sb] at 3.3492 Å and four equivalent bonds with the buckled square of four selenium at a distance of [Gd-Se1] 3.006 Å. Further, the buckled GdSe$_4$ is capped by another selenium[Se2] atom at a [Gd-Se2] distance of 2.9135 Å, forming the distorted tricapped trigonal prism\cite{Gebauer2021}. The coordination polyhedra of selenium can be described as an inverted monocapped buckled square of SeGd$_4$, as depicted in \hyperref[Fig. 1]{Fig. 1(f)}. The EDX (Energy Dispersive X-ray) pattern shown in \hyperref[Fig. 1]{Fig. 1}(b) confirms the homogenous distribution of constituent elements in the (1:1:1) ratio. The FESEM image shown in the upper inset of \hyperref[Fig. 1]{Fig. 1}(a) suggests the layered nature of material. 
\begin{figure}[t!]
\includegraphics[width=1.0 \columnwidth,angle=0,clip=true]{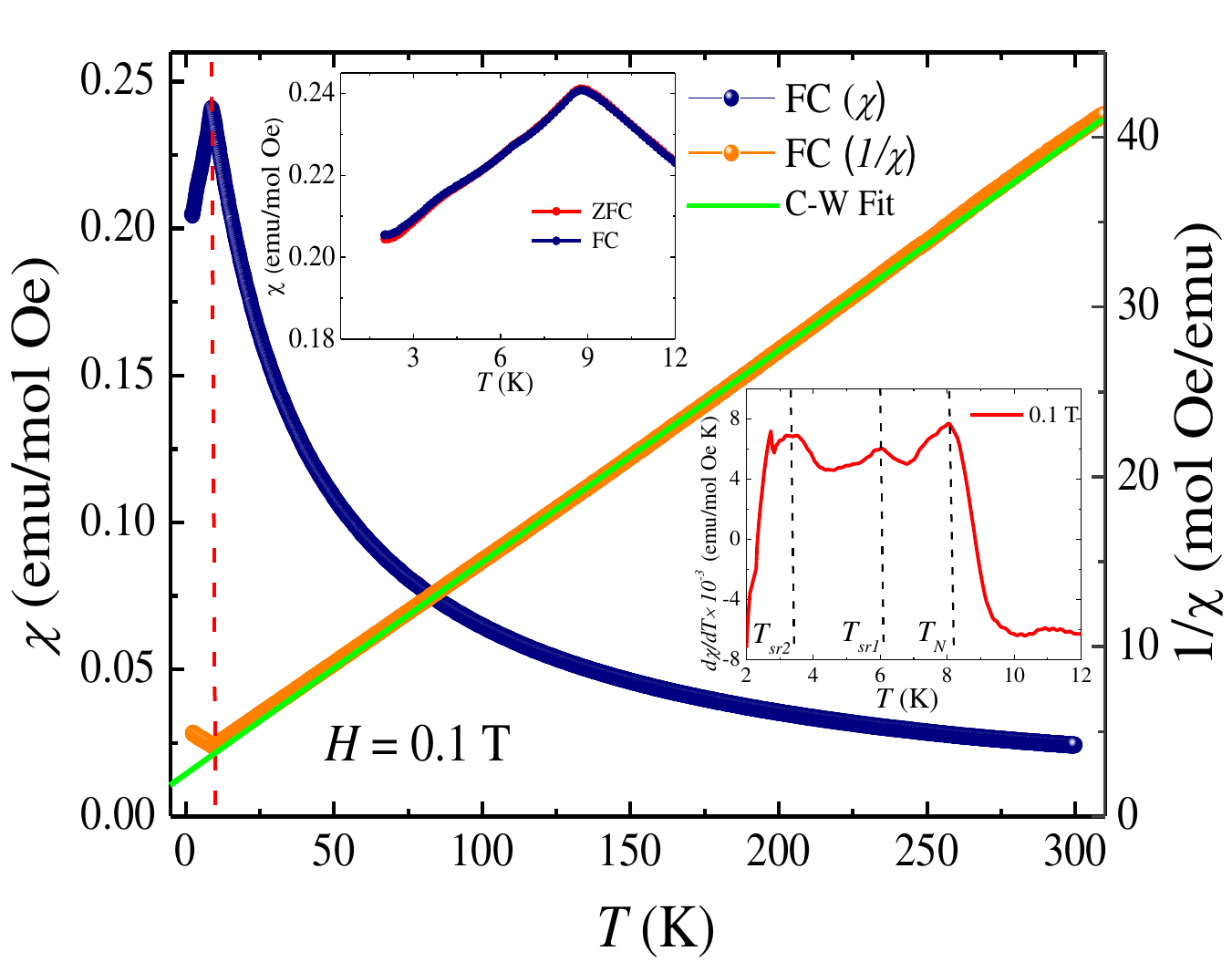}
\caption{(Color online) (a) Temperature dependence of the magnetic susceptibility ($\chi$) under Zero-Field cooled (ZFC) and Field cooled (FC) modes at an applied magnetic field of 0.1 T in the temperature range of 2-300 K (left scale). On the right scale, the reciprocal of magnetic susceptibility (1/$\chi$) for FC data as a function of temperature is presented. The green solid line indicates the Curie-Weiss fit in the temperature range of 10-300 K. The upper inset shows the expanded view of the $\chi$  vs T curve with no signatures of divergence under ZFC and FC protocols in low-temperature regions. The lower inset displays the derivative plot of magnetic susceptibility vs temperature at $H$ = 0.1 T.}
\label{Fig. 2}
\end{figure}

\begin{figure}[t!]
\includegraphics[width=1.05\columnwidth,angle=0,clip=true]{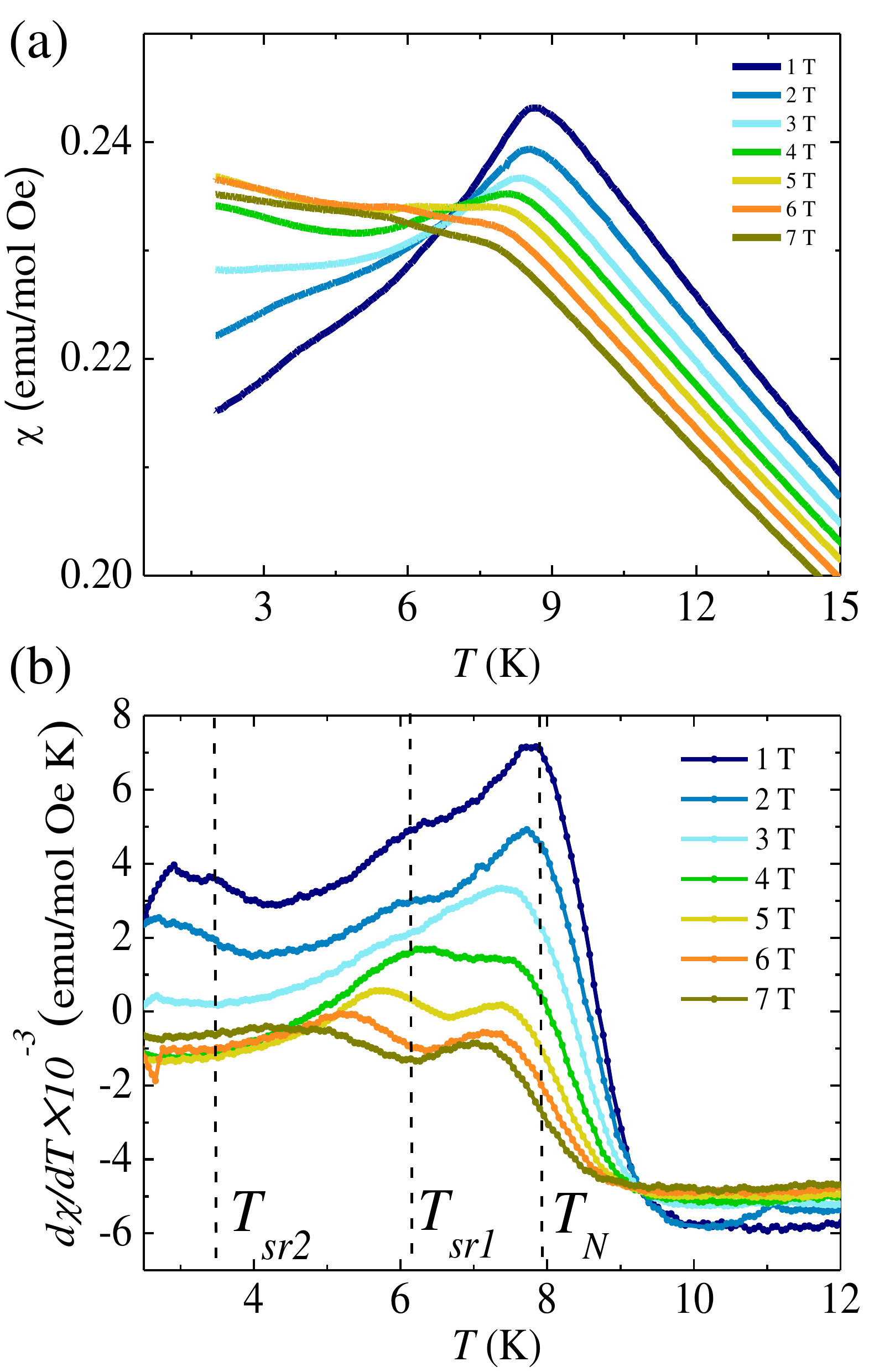}
\caption{(Color online) (a) Temperature-dependent DC magnetic susceptibility $\chi$(T)  at different applied magnetic fields ranging from 1 T to 7 T with an interval of $\Delta H = 1$ T in the temperature range of 2 - 15 K. (b) Derivative curve of magnetic susceptibility d$\chi$/dT as a function of temperature ($T$) ranging from 2 $-$ 12 K, displaying the two distinct spin-reorientation transition at different applied magnetic fields.}
\label{Fig. 3}
\end{figure}

\begin{table}[h]
\scriptsize\addtolength{\tabcolsep}{-1pt}
\caption{\label{table1}Refined structural and magnetic parameters of GdSbSe.}
\centering
\begin{ruledtabular}
\begin{tabular}{l c c c c c c}
GdSbSe \\
Space group: & $P4/nmm$ \\
Space group number: & 129\\
{\it a}(\r{A}): & 4.1340(2) \\
{\it c}(\r{A}): & 8.8870(6) \\
\hline

Atom & Site & x & y & z & Occu. & B$_{iso}$\\
Gd & 2c & 0.75 & 0.75 & 0.2968(3) & 1 & 0.78(9) \\
Sb & 2a & 0.75 & 0.25 & 0 & 1 & 0.76(9) \\
Se & 2c & 0.25 & 0.25 & 0.3748(5) & 1 & 0.85(1) \\

\hline

$T_N$ &&&& \multicolumn{3}{r}{8.6 $\pm$ 0.5 K}\\
$\Theta_P$ &&&& \multicolumn{3}{r}{-21.73 $\pm$ 0.03 K}   \\
$C$  &&&& \multicolumn{3}{r}{7.80$\pm$0.005 emuOe$^{-1}$mol$^{-1}$K$^{-1}$}   \\
$\mu_{eff}^{theo}$ &&&& \multicolumn{3}{r}{7.93$\mu_B$/Gd$^{+3}$}   \\
$\mu_{eff}^{calc}$ &&&& \multicolumn{3}{r}{7.89$\pm$0.01$\mu_B$/Gd$^{+3}$}   \\
$M_S$ at 2K &&&& \multicolumn{3}{r}{2.97$\pm$0.02 $\mu_B$/ f.u}   \\
$H_C$ at 2K &&&& \multicolumn{3}{r}{41$\pm$5 Oe}   \\

\end{tabular}
\end{ruledtabular}
\end{table}

\subsection{DC magnetic susceptibility}
Conventional magnetic characterization has been carried out through various measurement protocols to elucidate the magnetic properties of the compound. \hyperref[Fig. 2]{Fig. 2} shows the DC magnetic susceptibility as a function of temperature under the application of magnetic field ($H$ = 0.1 T) in both ZFC and FC protocols. Magnetic susceptibility reveals a sharp drop in magnetization below $T_N$ \(\approx\) 8.6 K, a typical signature observed for antiferromagnetic transitions. In addition, the absence of bifurcation in ZFC and FC curves eludes the possibility of magnetic irreversibility as shown in the inset of \hyperref[Fig. 2]{Fig. 2}. Further, to investigate the effect of external magnetic field on AFM interaction strength, we measured magnetic susceptibility at various external magnetic fields ranging from $H$ $=$ 1 T to 7 T. As displayed in  \hyperref[Fig. 3]{Fig. 3}(a), the antiferromagnetic transition becomes broader and shifts towards lower temperature with increasing field strength. This suggests the tendency of spin to attain a field-induced state. Interestingly, the antiferromagnetic transition is accompanied by two additional anomalies below Néel temperature which are discernible through the two sharp dips in the derivative plot of magnetic susceptibility d$\chi$/dT, as shown in the inset of \hyperref[Fig. 2]{Fig. 2}(a). The two anomalies below Néel temperature signal towards the possibility of spin-reorientation transition which has been reported in other Gd-based systems\cite{Ram2023, Sahu2021}. It is noteworthy that the anomalies below Néel temperature as shown by the black dotted lines, marked as  T$_\textit{sr1}$ and T$_\textit{sr2}$ exhibit strong field dependence and shifts towards lower temperatures at higher fields as evident through \hyperref[Fig. 3]{Fig. 3}(b). However, at field H $>$ 3 T, the magnitude of d$\chi$/dT increases below Néel temperature indicating the field-induced alignment of spins to another state. Further, for quantification of magnetic susceptibility results, the plot of inverse of magnetic susceptibility ($\chi^{-1}$) as a function of temperature (T) in temperature range of 2-300 K is presented. As shown in right scale of \hyperref[Fig. 2]{Fig. 2}, the curve is fitted with Curie-Weiss law using the following relation.

\begin{equation}\label{CW}
\chi (T) =  \frac{C}{(T-\theta_p)}
\end{equation}

The Curie constant $C$ is related to effective magnetic moments ($\mu_{eff}$) of ions present in the material through the relation, $C =  N_A\mu_{eff}^2/3k_B$, where $N_A$ represents Avogadro number, $k_B$ is the Boltzmann constant and, $\theta_p$ denotes the Weiss temperature of the material. The effective magnetic moment extracted from Curie-Weiss fit in temperature region 10 K- 300 K is 7.89 $\mu_B$/Gd$^{+3}$ which is consistent with the theoretical free ion value for Gd$^{+3}$ ion,  $\mu_{theo}$ = 7.93 $\mu_B$. The calculated value of $\theta_p$ from the fit is -21.73 K, here negative sign implies predominant AFM interactions among Gd$^{+3}$ spins.

\begin{figure}[t!]
\includegraphics[width=1.0 \columnwidth,angle=0,clip=true]{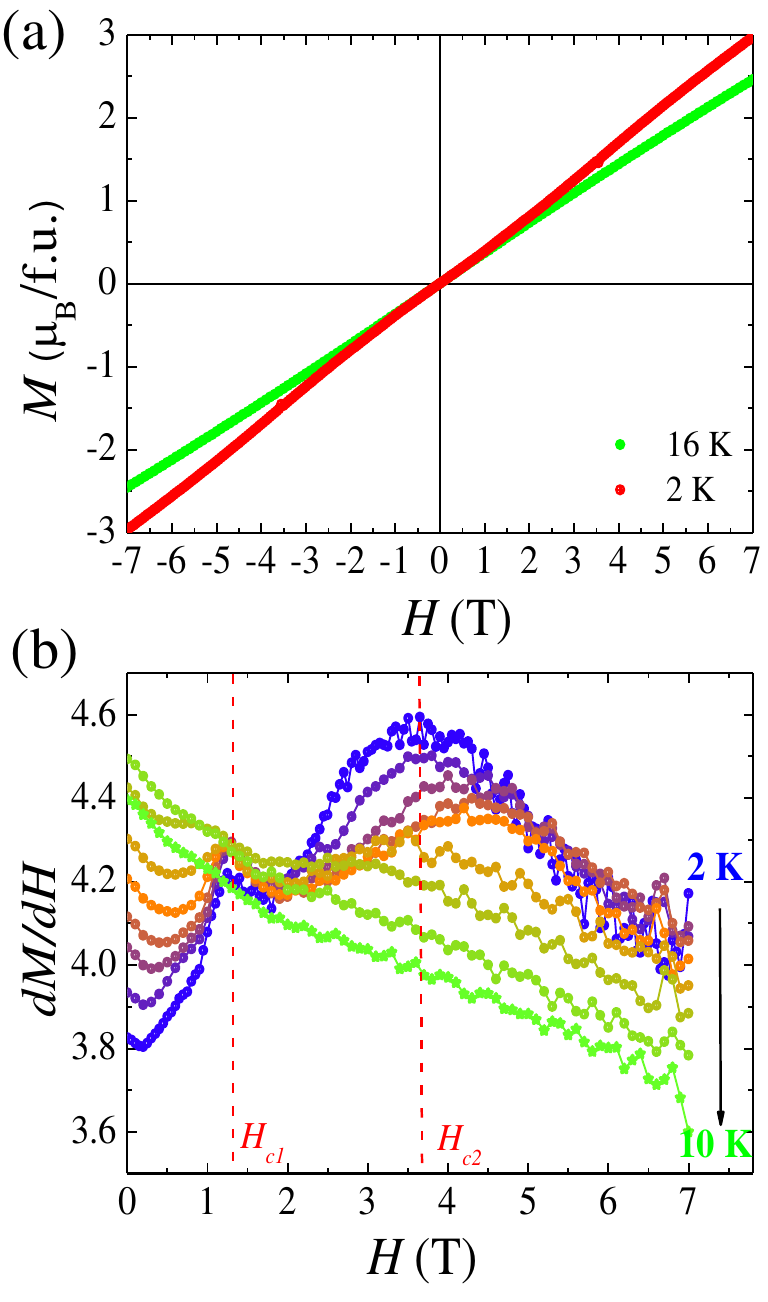}
\caption{(Color online) (a) Isothermal magnetization $M(H$) as a function of the magnetic field at 2 K and 16 K in the field range of -7 $T$ to +7 $T$. (b) The derivative plot of magnetization $dM/dH$ as a function of the applied magnetic field ($H$) at various temperatures across the magnetic transition from 2 K to 10 K, with the temperature interval of $\Delta T = 1$ K. The dashed red line represents the critical fields ($H_{C1}$ and $H_{C2}$) for the 2 K data.}
\label{Fig. 4}
\end{figure}

\begin{figure*}
\begin{center}
\includegraphics[width= 2.05\columnwidth,angle=0,clip=true]{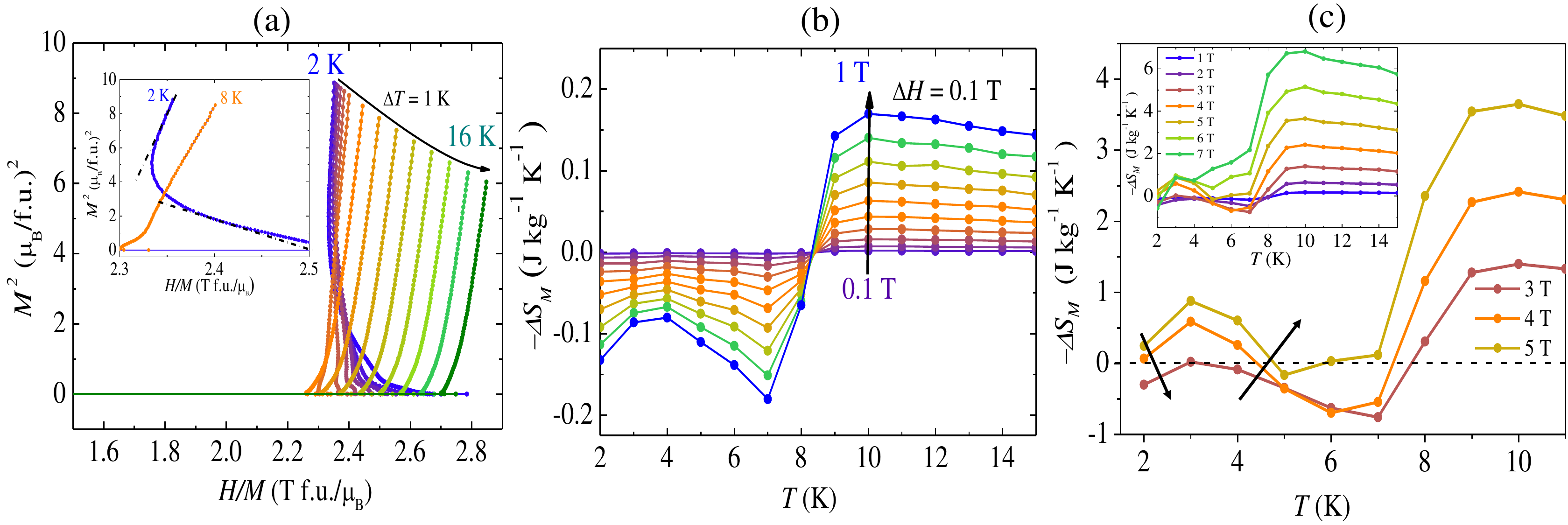}
\caption{(Color online) (a) Arrott plots (M$^2$ vs H/M) at different temperatures ranging from 2 K $-$ 16 K with temperature interval of $\Delta$ T = 1 K. The inset shows the expanded view of Arrott plots below and above Néel temperature, displaying the change in slope with field strength. (b) -$\Delta S_M$ in the low-field region $H$ = 0 T -1 T with an interval of $\Delta$ H = 0.1 T. (c) -$\Delta S_M$ at 3 T, 4 T, and 5 T in the temperature scale of 1.5 K - 11 K. The inset shows the temperature-dependent magnetic entropy change -$\Delta S_M$ in the high-field region $H$ = 1 T -7 T with an interval of $\Delta$ H = 1 T.}
\label{Fig. 5}
\end{center}
\end{figure*}

\subsection{Isothermal magnetization}

 To further investigate the nature of magnetic states at different temperatures ( 2 K and 16 K) with varying fields, we carried out isothermal $M(H)$ magnetization measurements around Néel temperature $T_N$ $\approx$ 8.6 K with an applied field of $\pm 7$ T. As shown in \hyperref[Fig. 4]{Fig. 4(a)}, $M(H)$ curves exhibit deviation from linearity below Néel temperature with no signatures of hysteresis, consistent with the AFM ground state below Néel temperature whereas it approaches linearity above T$_N$, implying the paramagnetic configuration of spins above transition temperature. Interestingly, the change in slope of virgin curves can be visualized in the field-dependent differential magnetization (d$M$/dH) plots as shown in \hyperref[Fig. 4]{Fig. 4(b)}. The d$M$/d$H$ curves display a peak followed by a broad shoulder below T$_N$ at two distinct critical fields H$_{c1}$  $\approx$ 1.26(3) T and  H$_{c2}$ $\approx$ 3.65(2) T at 2 K ,respectively. This behavior implies the possibility of metamagnetic transitions(MM) at two distinct fields. These MM transitions exhibit different dependence on the critical field as tabulated in Table \ref{table2}. The critical field corresponding to the first metamagnetic transition(MM1) does not exhibit significant change up to 5 K. In contrast, the critical field required for inducing MM2 increases with a temperature up to 5 K. This behavior is in close accordance with the previous antiferromagnetic systems displaying metamagnetic transitions\cite{Chakraborty2022}. Interestingly, we observed a reduction in both H$_{c1}$ and H$_{c2}$ around 7 K which completely disappeared at temperatures higher than 10 K. This behavior may arise due to the weakening of magnetic anisotropy energy at higher temperatures\cite{Muthuselvam2019}. It is noteworthy that due to the polycrystalline nature of GdSbSe, the anisotropic magnetic features merge, which is discernible through the two distinct metamagnetic transitions, possibly resulting from different spin alignments in specific directions. Further, the saturation magnetization achieved at the lowest temperature (2 K) and maximum field of 7 T is 2.97 $\mu_B$/f.u. , which is approximately 1/3 of the value of saturation magnetization M$_s$ $\approx$ 7.93 $\mu_B$/f.u expected for Gd$^{+3}$ ion. The significantly smaller moment value may be attributed to a distinct canted AFM spin arrangement with partial spin polarization in the field's direction. 
 \begin{table}[h]
\scriptsize\addtolength{\tabcolsep}{-0.65pt}
\caption{\label{table2} Extracted critical fields H$_{c1}$ and H$_{c2}$ at different temperatures from the derivative of isothermal magnetization data displayed in \hyperref[Fig. 4]{Fig. 4(b)}.}
\centering
\begin{tabular}{p{2.75 cm}p{3.75 cm}p{1.75 cm}}
\\
\hline
T (K) & $H$$_{c1}(T)$ & $H$$_{c2}(T)$ \\
\hline
{2} & 1.26(3) & 3.65(2) \\
{3} & 1.29(6) & 3.80(2) \\
{4} & 1.29(5) & 3.99(8) \\
{5} & 1.29(3) & 4.31(2) \\
{6} & 1.19(4) & 4.19(6) \\
{7} & 1.19(5) & 3.52(4) \\
{8} & 0.81(3) & 2.50(1) \\
{9} & 0.80(4) & 2.24(3) \\
\hline
\end{tabular}
\end{table}

\begin{figure}[t!]
\includegraphics[width=1.0\columnwidth,angle=0,clip=true]{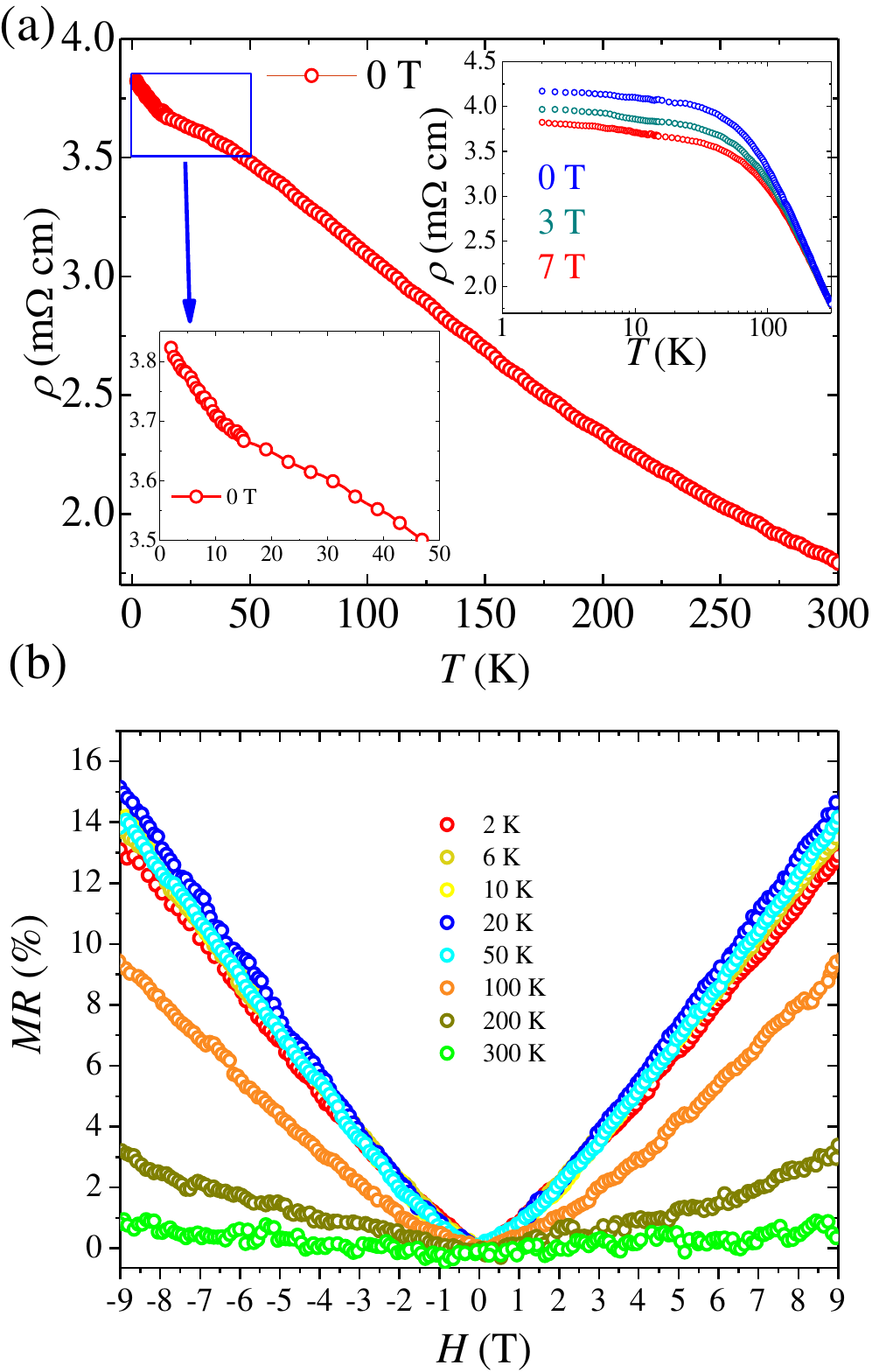}
 \caption{(Color online) (a) Temperature-dependent electrical resistivity of GdSbSe in the absence of magnetic field ( H = 0 T) in the temperature range of 2 K - 300 K. Lower inset shows the expanded view of resistivity displaying a sharp rise in close proximity of Néel temperature; upper inset represents the logarithmic temperature dependence of resistivity at 0 T, 3 T, and 7 T.(b) Magnetoresistance as a function of applied magnetic field at different temperature in the field range of -9 T to +9 T.}
\label{Fig. 6}
\end{figure}

\begin{figure}[t!]
\includegraphics[width=0.95\columnwidth,angle=0,clip=true]{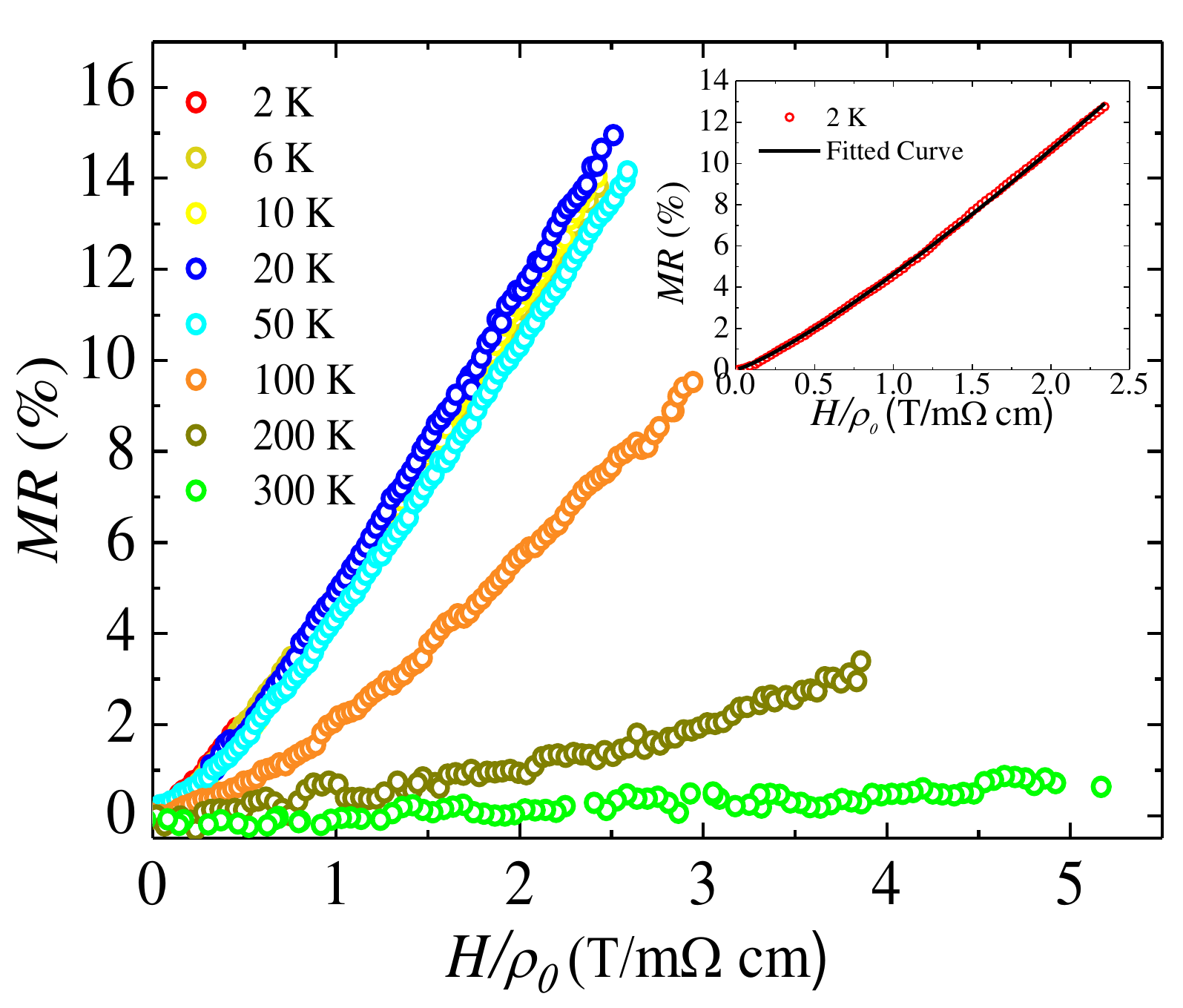}
 \caption{(Color online) (a) Magnetoresistance at different temperature scaled using Kohler's analysis; inset shows the power law (Eq-3) fitting of the experimental data at 2 K.}
\label{Fig. 7}
\end{figure}

\subsection{Magnetocaloric studies}

MCE (Magnetocaloric Effect) not only offers practical application advantages but also serves as an additional tool for understanding the magnetic interactions within a system. As established earlier, an enhanced magnetocaloric effect (MCE) is expected in systems displaying metamagnetic transitions. Double-phase transition systems are promising for MCE properties as they exhibit a significant change in magnetization, leading to a wide entropy change across the transition temperature\cite{Li2015, Sahu2021}. Therefore, we were motivated to extend our study to investigate the magnetocaloric properties of GdSbSe. Arrott plots ( M$^2$ vs H/M) were derived using the virgin curves observed at different temperatures ranging from 2K to 16 K in the field range H = 0 $-$ 7 T as shown in \hyperref[Fig. 5]{Fig. 5(a)}. According to Banerjee criteria, the materials exhibiting positive and negative slopes display second-order and first-order magnetic phase transitions,respectively\cite{Banerjee1964}. As demonstrated in  \hyperref[Fig. 5]{Fig. 5(a)}, Arrott curves exhibit a negative slope below the Néel temperature in the low-field region, which transitions to a positive slope above the Néel temperature which indicates a first-order phase transition from the antiferromagnetic (AFM) to the paramagnetic (PM) state\cite{Mishra2024a, Mishra2024c}. Interestingly, the slope also changes from negative to positive in the high-field region, implying a field-induced metamagnetic transition. According to Maxwell's relation, the change in magnetic entropy, $\Delta S_M$ associated with the magnetic transition can be derived using the thermodynamic relation by integrating a range of applied fields.

\begin{equation}\label{Entropy2}
\Delta S (T, H_0)=  \int^{H_0}_0 \left[ \frac{\partial M (T, H)}{\partial T} \right]_{H} dH
\end{equation}
 \hyperref[Fig. 5]{Figure. 5(b)} shows the temperature dependence of  -$\Delta S_M$ as a function of temperature in the low-field region ( H$<$ 1 T). The crossover from negative to positive values in $\Delta S_M$ curves around ordering temperature (T$_N$ $\approx 8.6$ K), indicates a transition from conventional magnetocaloric effect (MCE) to inverse magnetocaloric effect (IMCE), which is commonly observed for AFM systems\cite{Biswas2013}. It is noteworthy that $\Delta S_M$ remains positive below ordering temperature up to a maximum field of 2 T, suggesting the robustness of the AFM state below Néel temperature. Interestingly, we observed two subsequent crossovers in entropy curves around 6 K and 3 K  for H = 4T, respectively, as depicted through the black arrow in \hyperref[Fig. 5]{Fig. 5(c)}. The transitions are concomitant with the spin-reorientation behavior established through DC susceptibility measurements. Moreover, the behavior of -$\Delta S_M$ under a high magnetic field is also investigated. As shown in the inset of \hyperref[Fig. 5]{Fig. 5(c)}, $\Delta S_M$ attains a negative value of  6.77 JKg$^{-1}$K$^{-1}$ for the maximum field of 7 T. The overall negative value of $\Delta S_M$ for H $\geq5T$ suggests the field-induced ferromagnetic state\cite{Mishra2024b, Kumar2020}. 

 \subsection{Transport and magnetotransport studies}

\begin{figure*}
\begin{center}
\includegraphics[width= 1.95\columnwidth,angle=0,clip=true]{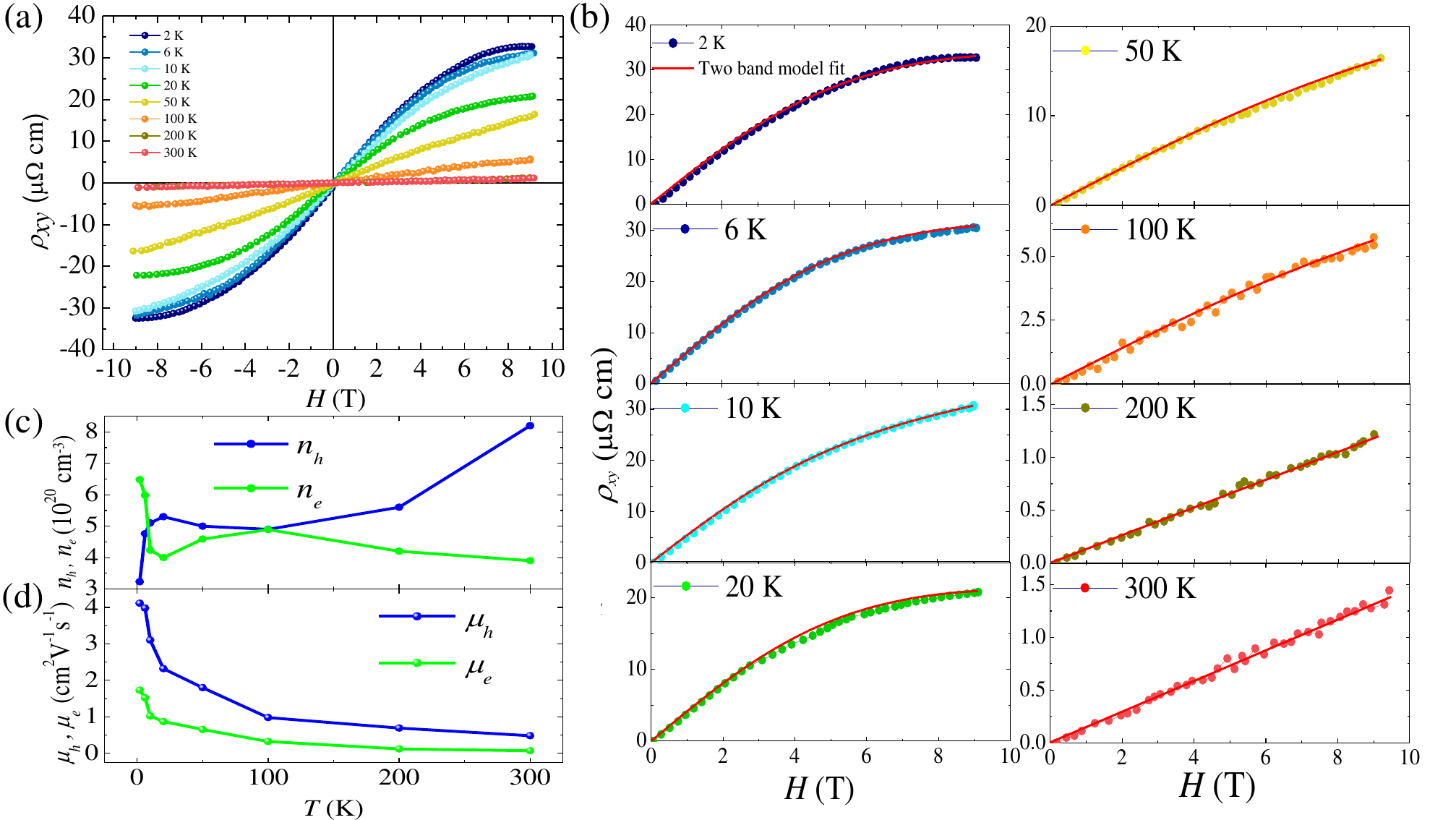}
\caption{(Color online)(a) Hall resistivity measured at various temperatures from 2 K to 300 K in the field range of $-$ 9 T to $+$ 9 T.(b) Hall resistivity data fitted using two-band model, red line shows the fitting at different temperatures. (c)  and (d) The variation of carrier density (\( n_h \), \( n_e \)) and the mobility (\(\mu_h \) ,\(\mu_e \)) with temperature derived from the two-band model fit.}
\label{Fig. 8}
\end{center}
\end{figure*}
To gain further insights into the correlation between magnetism and transport properties of the charge carriers, we investigated the transport and magneto-transport behavior of GdSbSe. \hyperref[Fig. 6]{Fig. 6(a)} represents the temperature-dependent longitudinal resistivity ($\rho(T)$) at zero applied magnetic field. The $\rho(T)$ exhibits semimetallic behavior with decreasing temperature followed by a sudden increase close to Néel temperature. The cusp-like anomaly possibly indicates the zero-band gap semimetallic behavior of GdSbSe. This kind of cusp-like anomaly accompanied by a sharp increase in resistivity in the proximity of Néel temperature has been observed in GdPtBi\cite{Schindler2020}. To probe the effect of field strength, the $\rho(T)$ data were recorded at different applied fields and plotted at a logarithmic temperature scale, as shown in the upper inset of \hyperref[Fig. 6]{Fig. 6(a)}. The $\rho(T)$ increases with increasing field strength consistent with the positive magnetoresistance observed in GdSbSe. Intriguingly, other rare-earth analogs of this series do not exhibit the cusp-like anomalies in resistivity behavior such as LaSbTe displays metallic nature and Nd/Sm(SbTe) displays non-metallic temperature dependence\cite{Singha2017a}\cite{Pandey2020}\cite{Pandey2021}. 
Further, longitudinal magnetoresistance with varying magnetic fields up to 9 T is shown in \hyperref[Fig. 4]{Fig. 6(b)}. MR(Magnetoresistance) is defined as $\frac{\rho(H)-\rho(0)}{\rho(0)} \times 100$, where $\rho(H)$ and $\rho(0)$ are resistivity in the presence and absence of applied magnetic field, respectively. It can be observed that a positive non-saturating magneto-resistance behavior with an MR of $\approx$ 13\% is observed at 2 K at the maximum applied magnetic field of 9 T. Interestingly, MR attains a maximum value of $\approx$ 15\% at 20 K with the maximum applied magnetic field of 9 T which decreases with increasing temperature. The obtained MR\% is significantly higher than the previously reported selenium analogs of LnSbSe series\cite{Pandey2022}. According to the classical MR model, magnetoresistance curves display quadratic field dependence MR $\propto H^2$  for normal metals and semiconductors\cite{Rossiter2006}. To deeply investigate the presence of crossover from quadratic to linear dependence in magnetoresistance curves, we employed Kohler's scaling analysis and fitted the data at 2 K using the power-law equation-
\begin{equation}\label{Entropy}
MR = \alpha \left( \left( \frac{H}{\rho_0} \right)^m \right)
\end{equation}

\hyperref[Fig. 7]{Fig. 7} represents the Kohler's plots at different temperatures derived from the magnetoresistance data. The inset of \hyperref[Fig. 7]{Fig. 7} shows the fitted curve at 2 K. The extracted parameter from the fit yields $\alpha$ = 4.661 m$\ohm$cm/T and m = 1.20. which suggests the deviation from quadratic dependence of the field of MR at low temperatures. This observation strengthens the possibility of non-trivial characteristics such as Dirac-like dispersion in this kind of material, which has been established earlier in other isostructural analogs of this series\cite{Yuan2024, Gao2022, Yue2020}. According to Kohler's rule, magnetoresistance curves for materials with a single type of charge carrier can be scaled into a single curve due to the uniform scattering time across the Fermi surface. Furthermore, Kohler's plots exhibit a significant deviation from Kohler's rule, as the MR curves do not fall on a single curve\cite{Singh2024}. The deviation from Kohler's plot suggests the existence of more than one type of charge carrier in the material, consistent with the previously explored isostructural analogs of this series of compounds\cite{Pandey2022}. To elucidate the presence of multiple types of charge carriers, we measured Hall resistivity at different temperatures as discussed in the section below.


\subsection{Hall resistivity studies}
Further, we performed Hall measurements to determine the nature and density of charge carriers in the field range of $\pm  9$ T  at different temperatures. As displayed in \hyperref[Fig. 7]{Fig. 8(a)}, Hall resistivity as a function of field is plotted at different temperatures. Hall resistivity exhibits an almost positive linear dependence with the field at room temperature, indicating the dominance of hole-type charge carriers at room temperature. However, at low temperatures, a non-linearity in Hall resistivity curves starts to manifest which indicates the presence of more than one type of charge carriers in GdSbSe and is consistent with our previous observations from magnetoresistance data. These results signal the presence of electron and hole type of charge carriers and have been reported previously in other selenium analogs such as ZrSiSe\cite{Chen2020} and LaSbSe\cite{Pandey2022}. In contrast with the previous reports on GdSbTe, stating the presence of electron-type dominance in Hall resistivity measurements\cite{Gebauer2021}, we observed hole types of dominance at room temperature. This behavior strengthens the possibility of change in Fermi energy via selenium substitution through varying unit cell dimensions. Further, for the quantitative analysis of the density and mobility of charge carriers, we followed the well-known two-band model. The Hall resistivity in the two-band model is given by:

\begin{equation}\label{Entropy}
\rho_{xy} = \frac{H}{e} \frac{n_h \mu_h^2 - n_e \mu_e^2 + (n_h - n_e) \mu_e^2 \mu_h^2 H^2}{(n_e \mu_e + n_h \mu_h)^2 + (n_h - n_e)^2 \mu_e^2 \mu_h^2 H^2}
\end{equation}

where \( n_e \) and \( n_h \) denote the carrier density of electrons and holes, respectively, and  \(\mu_e \) and \( \mu_h \) represent the mobility of electrons and holes, respectively. \hyperref[Fig. 8]{Fig. 8(b)} shows the fitted curves of the two-band model at different temperatures in the field range of 0 - 9 T. The parameters extracted from the fitting at 2 K are \( n_h \) =$6.56 \times 10^{20}$ cm$^{-3}$ \( n_e \)= $3.18 \times 10^{20}$ cm$^{-3}$, significantly larger than other Dirac semimetals such as Cd$_3$As$_2$ (10$^{16}$-10$^{17}$ and Na$_3$Bi (10$^{17}$-10$^{18}$) \cite{Feng2017, Xiong2015, He2014} but comparable to the ZrSiS type nodal line semimetals\cite{Hu2016}.The obtained values of hole and electron mobilities at 2K are \( \mu_h \) = 411 cm$^2$V$^{-1}$ s$^{-1}$ and \( \mu_e \) = 173 cm$^2$V$^{-1}$ s$^{-1}$ comparatively lower than the carrier mobilities observed in isostructural nodal-line semimetals\cite{Sankar2017} but consistent with the carrier mobilities reported for non-magnetic analog LaSbSe.  It may be noted that the presence of electrons as charge carriers manifests the possibility of tuned Fermi level with selenium substitution. However, theoretical analysis can shed more light on the evolution of electron and hole pockets due to Fermi surface reconstruction as a function of temperatures. The electron density decreases monotonically with increasing temperatures whereas the density of holes increases with temperature as discernible from \hyperref[Fig. 8]{Fig. 8(c)}. In addition, the mobility of electrons and holes decreases with increasing temperature as illustrated in \hyperref[Fig. 8]{Fig. 8(d)}. These observations signal the possibility of GdSbSe as a Dirac nodal-line semimetal with non-trivial characteristics as its other isostructural compounds.

\subsection{Specific heat}

Specific heat measurements have been performed at various applied magnetic fields to corroborate the findings of magnetic transitions in our previous studies. \hyperref[Fig. 8]{Fig. 9(a)} shows the heat capacity of GdSbSe measured in the temperature range of 2-100 K. A broad $\lambda$- peak anomaly close to T$_N$ is observed, consistent with the AFM transition evident from the magnetic measurements. Interestingly, a close expanded view of the peak below T$_N$ reveals the splitting of peaks in two parts as shown in the upper inset of  \hyperref[Fig. 8]{Fig. 9(a)}. This behavior suggests the reorientation of the spin process in the compound and has been observed in other Gd-based materials\cite{Ram2023}. Furthermore, the evolution of the spin reorientation process has been investigated by observing the data in different applied fields through $C/T$ vs $T$ curves as depicted in the inset of \hyperref[Fig. 8]{Fig. 9(b)}.The concomitant spin reorientation transition, identified from DC susceptibility measurements shifts to lower temperatures with increasing field strength as shown by the dashed lines in the inset of \hyperref[Fig. 8]{Fig. 9(b)}. The temperature dependence of specific heat ($C(T)$), in the low-temperature region, can be expressed as-

\begin{equation}\label{Debye}
C= \gamma T + \beta T^3 + C_m
\end{equation}
where $\gamma$ is the Sommerfeld coefficient and $\beta$ represents the phonon contribution coefficient, and $C_m$ is the magnetic contribution to specific heat. Considering the negligible magnetic contribution at high temperatures, a linear $T^2$ dependency is discernible from the fit of $C/T$ vs $T^2$ plot within the temperature range of 15 K$\leq$T$\leq$25 K using the equation \begin{equation}\label{Debye} C/T= \gamma  + \beta T^2 \end{equation} as shown by the red solid line in the inset of \hyperref[Fig. 8]{Fig. 9(a)}.

\begin{figure}[t!]
\includegraphics[width=1.0\columnwidth,angle=0,clip=true]{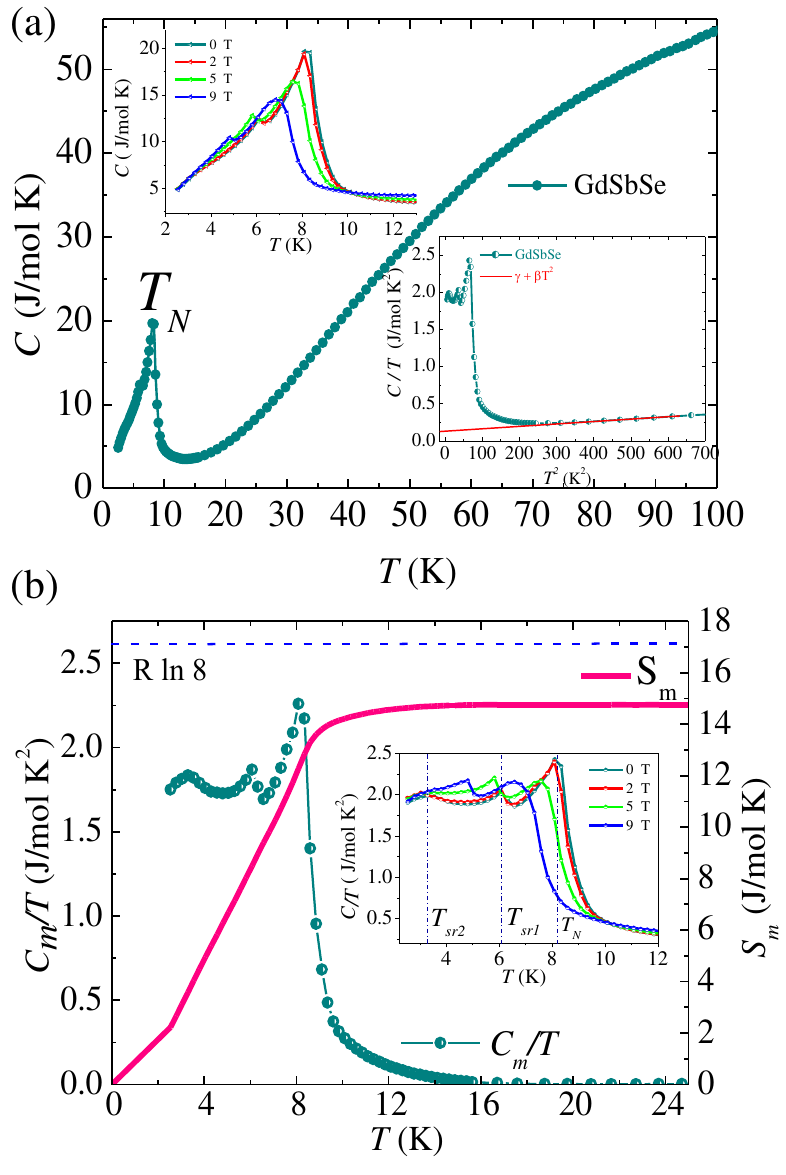}
 \caption{(Color online) (a) Temperature-dependent heat capacity $C$(T) of GdSbSe in the temperature range of 2 K $-$ 100 K; upper inset displays the ($C$) vs T at the different applied magnetic field in the temperature range of 2 - 13 K. the lower inset shows C/T vs T$^2$ in the low-temperature range, the red line shows the fitting of data using the equation described in the text.(b) {C$_m$}/T vs T in the low-temperature range of 2 K-25 K on the left scale and the magnetic entropy S$_m$ as a function of temperature; $C$/T vs T at different magnetic fields in the temperature range of 2 K - 12 K.}
\label{Fig. 9}
\end{figure}
The value of extracted parameters from the linear fit of the above-mentioned equation is $\gamma$ and $\beta$ is 152(5) mJ/ mol K$^2$ and 0.284(3) mJ/mol K$^4$, respectively. The relatively large value of the Sommerfeld coefficient compared to other isostructural antiferromagnetic (41 mJ/mol K$^2$ for CeSbTe),(8.6 mJ/molK$^2$ mol for CeSbSe\cite{Chen2017}) and 2.19 mJ/ mol K$^2$ for non-magnetic LaSbSe\cite{Pandey2022} suggests the enhanced electronic correlations in the material which is often associated with heavy Fermionic behavior in the system \cite{Yang2020}. From the fit, the value of the Debye temperature $\theta_D$ has been obtained using the formula $\theta_D= (12\pi^4NR/5\beta)^{1/3}$, where N = 3 and  R represents universal gas constant. The estimated value of $\theta_D$ is 273 K, significantly higher than the tellurium analog GdSbTe, implying stronger interatomic interactions in GdSbSe.
For further extraction of magnetic entropy, the magnetic contribution is deduced by subtracting the electronic and phononic contributions from the specific heat data. The magnetic entropy is calculated by integrating C$_m$/T over T using the relation $S_m = \int_0^T \frac{C_m(T)}{T}dT$ up to 25 K as shown in the right scale of \hyperref[Fig. 8]{Fig. 9(b)}, the magnetic entropy increases sharply up to 9 K and then saturates with a further increase in temperature. It may be noted that the entropy between 0 K and 2 K is estimated through the linear extrapolation of the entropy curve, discernible through the entropy curve. The maximum attained value of magnetic entropy is 14.78 JK$^{-1}$mol$^{-1}$ which is reasonably close to the theoretically estimated magnetic entropy S = Rln(2S+1) = 17.28 JK$^{-1}$mol$^{-1}$ for S = 7/2.


\section{CONCLUSIONS}
In conclusion, we have investigated the magnetic, transport, and magneto-transport properties of GdSbSe. The temperature-dependent magnetic susceptibility measurements reveal antiferromagnetic ordering with Néel temperature T$_N$ = 8.6 K. The $\lambda$-shaped anomaly is strongly manifested through specific heat measurements, alongside signatures of spin reorientation consistent with magnetic measurements. Isothermal magnetization $M(H)$ measurements reveal the presence of metamagnetic transition (MM) below T$_N$ at two critical fields $H$$_{c1}$ = 1.26 T and $H$$_{c2}$ = 3.65 T. Magnetocaloric studies (MCE) in GdSbSe unveiled the signatures of field-induced metamagnetic transition via crossover from the inverse magnetocaloric effect (IMCE) to the conventional magnetocaloric effect. Further, transport studies establish GdSbSe as a semi-metallic compound that is in close correlation with magnetic observation in the proximity of Néel temperature. This correlation suggests a strong coupling of the electronic and magnetic properties of the system. Further, magneto-transport studies indicate non-quadratic field dependence, indicating the presence of Dirac-like dispersion at the Fermi level. The manifestation of two types of charge carriers through violation of Kohler's law and non-linearity in Hall resistivity measurements further supports the possibility of a multiband nature in GdSbSe.The aforementioned observation strongly suggests non-trivial characteristic features in GdSbSe and suggests various topological surface states. However, theoretical band structure calculations and ARPES measurements can shed more light on the topological features.

\section*{ACKNOWLEDGMENTS}
The authors acknowledge the Central Research Facility (CRF), IIT Delhi for experimental facilities. We thank the MPMS3 facility of the Physics department, IIT Delhi for the magnetic measurements. AG acknowledge the Council of Scientific $\&$ Industrial Research (CSIR), [09/0086(12689)/2021-EMR-I] India for fellowship. SB and AS thank Sheikh Saqr Laboratory at Jawaharlal Nehru Centre for Advanced Scientific Research for all the transport and specific heat measurements. SB thanks CSIR for the fellowship. AKG expresses gratitude to SERB for the financial assistance from Project Number [CRG/2022/000178], Government of India.

\bibliographystyle{apsrev4-2}

\begin{thebibliography}{57}%
\makeatletter
\providecommand \@ifxundefined [1]{%
 \@ifx{#1\undefined}
}%
\providecommand \@ifnum [1]{%
 \ifnum #1\expandafter \@firstoftwo
 \else \expandafter \@secondoftwo
 \fi
}%
\providecommand \@ifx [1]{%
 \ifx #1\expandafter \@firstoftwo
 \else \expandafter \@secondoftwo
 \fi
}%
\providecommand \natexlab [1]{#1}%
\providecommand \enquote  [1]{``#1''}%
\providecommand \bibnamefont  [1]{#1}%
\providecommand \bibfnamefont [1]{#1}%
\providecommand \citenamefont [1]{#1}%
\providecommand \href@noop [0]{\@secondoftwo}%
\providecommand \href [0]{\begingroup \@sanitize@url \@href}%
\providecommand \@href[1]{\@@startlink{#1}\@@href}%
\providecommand \@@href[1]{\endgroup#1\@@endlink}%
\providecommand \@sanitize@url [0]{\catcode `\\12\catcode `\$12\catcode `\&12\catcode `\#12\catcode `\^12\catcode `\_12\catcode `\%12\relax}%
\providecommand \@@startlink[1]{}%
\providecommand \@@endlink[0]{}%
\providecommand \url  [0]{\begingroup\@sanitize@url \@url }%
\providecommand \@url [1]{\endgroup\@href {#1}{\urlprefix }}%
\providecommand \urlprefix  [0]{URL }%
\providecommand \Eprint [0]{\href }%
\providecommand \doibase [0]{https://doi.org/}%
\providecommand \selectlanguage [0]{\@gobble}%
\providecommand \bibinfo  [0]{\@secondoftwo}%
\providecommand \bibfield  [0]{\@secondoftwo}%
\providecommand \translation [1]{[#1]}%
\providecommand \BibitemOpen [0]{}%
\providecommand \bibitemStop [0]{}%
\providecommand \bibitemNoStop [0]{.\EOS\space}%
\providecommand \EOS [0]{\spacefactor3000\relax}%
\providecommand \BibitemShut  [1]{\csname bibitem#1\endcsname}%
\let\auto@bib@innerbib\@empty
\bibitem [{\citenamefont {Hasan}\ and\ \citenamefont {Kane}(2010)}]{Hasan2010}%
  \BibitemOpen
  \bibfield  {author} {\bibinfo {author} {\bibfnamefont {M.~Z.}\ \bibnamefont {Hasan}}\ and\ \bibinfo {author} {\bibfnamefont {C.~L.}\ \bibnamefont {Kane}},\ }\href {https://doi.org/10.1103/RevModPhys.82.3045} {\bibfield  {journal} {\bibinfo  {journal} {Rev. Mod. Phys.}\ }\textbf {\bibinfo {volume} {82}},\ \bibinfo {pages} {3045} (\bibinfo {year} {2010})}\BibitemShut {NoStop}%
\bibitem [{\citenamefont {Fu}(2011)}]{Fu2011}%
  \BibitemOpen
  \bibfield  {author} {\bibinfo {author} {\bibfnamefont {L.}~\bibnamefont {Fu}},\ }\href {https://doi.org/10.1103/PhysRevLett.106.106802} {\bibfield  {journal} {\bibinfo  {journal} {Phys. Rev. Lett.}\ }\textbf {\bibinfo {volume} {106}},\ \bibinfo {pages} {106802} (\bibinfo {year} {2011})}\BibitemShut {NoStop}%
\bibitem [{\citenamefont {Burkov}(2016)}]{burkov2016}%
  \BibitemOpen
  \bibfield  {author} {\bibinfo {author} {\bibfnamefont {A.}~\bibnamefont {Burkov}},\ }\href {https://doi.org/https://doi.org/10.1038/nmat4788} {\bibfield  {journal} {\bibinfo  {journal} {Nat. Mater.}\ }\textbf {\bibinfo {volume} {15}},\ \bibinfo {pages} {1145} (\bibinfo {year} {2016})}\BibitemShut {NoStop}%
\bibitem [{\citenamefont {Lv}\ \emph {et~al.}(2021)\citenamefont {Lv}, \citenamefont {Qian},\ and\ \citenamefont {Ding}}]{Lv2021}%
  \BibitemOpen
  \bibfield  {author} {\bibinfo {author} {\bibfnamefont {B.~Q.}\ \bibnamefont {Lv}}, \bibinfo {author} {\bibfnamefont {T.}~\bibnamefont {Qian}},\ and\ \bibinfo {author} {\bibfnamefont {H.}~\bibnamefont {Ding}},\ }\href {https://doi.org/10.1103/RevModPhys.93.025002} {\bibfield  {journal} {\bibinfo  {journal} {Rev. Mod. Phys.}\ }\textbf {\bibinfo {volume} {93}},\ \bibinfo {pages} {025002} (\bibinfo {year} {2021})}\BibitemShut {NoStop}%
\bibitem [{\citenamefont {Huang}\ \emph {et~al.}(2015)\citenamefont {Huang}, \citenamefont {Zhao}, \citenamefont {Long}, \citenamefont {Wang}, \citenamefont {Chen}, \citenamefont {Yang}, \citenamefont {Liang}, \citenamefont {Xue}, \citenamefont {Weng}, \citenamefont {Fang}, \citenamefont {Dai},\ and\ \citenamefont {Chen}}]{Huang2015}%
  \BibitemOpen
  \bibfield  {author} {\bibinfo {author} {\bibfnamefont {X.}~\bibnamefont {Huang}}, \bibinfo {author} {\bibfnamefont {L.}~\bibnamefont {Zhao}}, \bibinfo {author} {\bibfnamefont {Y.}~\bibnamefont {Long}}, \bibinfo {author} {\bibfnamefont {P.}~\bibnamefont {Wang}}, \bibinfo {author} {\bibfnamefont {D.}~\bibnamefont {Chen}}, \bibinfo {author} {\bibfnamefont {Z.}~\bibnamefont {Yang}}, \bibinfo {author} {\bibfnamefont {H.}~\bibnamefont {Liang}}, \bibinfo {author} {\bibfnamefont {M.}~\bibnamefont {Xue}}, \bibinfo {author} {\bibfnamefont {H.}~\bibnamefont {Weng}}, \bibinfo {author} {\bibfnamefont {Z.}~\bibnamefont {Fang}}, \bibinfo {author} {\bibfnamefont {X.}~\bibnamefont {Dai}},\ and\ \bibinfo {author} {\bibfnamefont {G.}~\bibnamefont {Chen}},\ }\href {https://doi.org/10.1103/PhysRevX.5.031023} {\bibfield  {journal} {\bibinfo  {journal} {Phys. Rev. X}\ }\textbf {\bibinfo {volume} {5}},\ \bibinfo {pages} {031023} (\bibinfo {year} {2015})}\BibitemShut {NoStop}%
\bibitem [{\citenamefont {Armitage}\ \emph {et~al.}(2018)\citenamefont {Armitage}, \citenamefont {Mele},\ and\ \citenamefont {Vishwanath}}]{Armitage2018}%
  \BibitemOpen
  \bibfield  {author} {\bibinfo {author} {\bibfnamefont {N.~P.}\ \bibnamefont {Armitage}}, \bibinfo {author} {\bibfnamefont {E.~J.}\ \bibnamefont {Mele}},\ and\ \bibinfo {author} {\bibfnamefont {A.}~\bibnamefont {Vishwanath}},\ }\href {https://doi.org/10.1103/RevModPhys.90.015001} {\bibfield  {journal} {\bibinfo  {journal} {Rev. Mod. Phys.}\ }\textbf {\bibinfo {volume} {90}},\ \bibinfo {pages} {015001} (\bibinfo {year} {2018})}\BibitemShut {NoStop}%
\bibitem [{\citenamefont {Young}\ \emph {et~al.}(2012)\citenamefont {Young}, \citenamefont {Zaheer}, \citenamefont {Teo}, \citenamefont {Kane}, \citenamefont {Mele},\ and\ \citenamefont {Rappe}}]{Young2012}%
  \BibitemOpen
  \bibfield  {author} {\bibinfo {author} {\bibfnamefont {S.~M.}\ \bibnamefont {Young}}, \bibinfo {author} {\bibfnamefont {S.}~\bibnamefont {Zaheer}}, \bibinfo {author} {\bibfnamefont {J.~C.~Y.}\ \bibnamefont {Teo}}, \bibinfo {author} {\bibfnamefont {C.~L.}\ \bibnamefont {Kane}}, \bibinfo {author} {\bibfnamefont {E.~J.}\ \bibnamefont {Mele}},\ and\ \bibinfo {author} {\bibfnamefont {A.~M.}\ \bibnamefont {Rappe}},\ }\href {https://doi.org/10.1103/PhysRevLett.108.140405} {\bibfield  {journal} {\bibinfo  {journal} {Phys. Rev. Lett.}\ }\textbf {\bibinfo {volume} {108}},\ \bibinfo {pages} {140405} (\bibinfo {year} {2012})}\BibitemShut {NoStop}%
\bibitem [{\citenamefont {Burkov}\ \emph {et~al.}(2011)\citenamefont {Burkov}, \citenamefont {Hook},\ and\ \citenamefont {Balents}}]{Burkov2011}%
  \BibitemOpen
  \bibfield  {author} {\bibinfo {author} {\bibfnamefont {A.~A.}\ \bibnamefont {Burkov}}, \bibinfo {author} {\bibfnamefont {M.~D.}\ \bibnamefont {Hook}},\ and\ \bibinfo {author} {\bibfnamefont {L.}~\bibnamefont {Balents}},\ }\href {https://doi.org/10.1103/PhysRevB.84.235126} {\bibfield  {journal} {\bibinfo  {journal} {Phys. Rev. B}\ }\textbf {\bibinfo {volume} {84}},\ \bibinfo {pages} {235126} (\bibinfo {year} {2011})}\BibitemShut {NoStop}%
\bibitem [{\citenamefont {Fang}\ \emph {et~al.}(2015)\citenamefont {Fang}, \citenamefont {Chen}, \citenamefont {Kee},\ and\ \citenamefont {Fu}}]{Fang2015}%
  \BibitemOpen
  \bibfield  {author} {\bibinfo {author} {\bibfnamefont {C.}~\bibnamefont {Fang}}, \bibinfo {author} {\bibfnamefont {Y.}~\bibnamefont {Chen}}, \bibinfo {author} {\bibfnamefont {H.-Y.}\ \bibnamefont {Kee}},\ and\ \bibinfo {author} {\bibfnamefont {L.}~\bibnamefont {Fu}},\ }\href {https://doi.org/10.1103/PhysRevB.92.081201} {\bibfield  {journal} {\bibinfo  {journal} {Phys. Rev. B}\ }\textbf {\bibinfo {volume} {92}},\ \bibinfo {pages} {081201} (\bibinfo {year} {2015})}\BibitemShut {NoStop}%
\bibitem [{\citenamefont {Klemenz}\ \emph {et~al.}(2019)\citenamefont {Klemenz}, \citenamefont {Lei},\ and\ \citenamefont {Schoop}}]{Klemenz2019}%
  \BibitemOpen
  \bibfield  {author} {\bibinfo {author} {\bibfnamefont {S.}~\bibnamefont {Klemenz}}, \bibinfo {author} {\bibfnamefont {S.}~\bibnamefont {Lei}},\ and\ \bibinfo {author} {\bibfnamefont {L.~M.}\ \bibnamefont {Schoop}},\ }\href {https://doi.org/10.1146/annurev-matsci-070218-010114} {\bibfield  {journal} {\bibinfo  {journal} {Annu. Rev. Mater. Res.}\ }\textbf {\bibinfo {volume} {49}},\ \bibinfo {pages} {185} (\bibinfo {year} {2019})}\BibitemShut {NoStop}%
\bibitem [{\citenamefont {Klemenz}\ \emph {et~al.}(2020)\citenamefont {Klemenz}, \citenamefont {Hay}, \citenamefont {Teicher}, \citenamefont {Topp}, \citenamefont {Cano},\ and\ \citenamefont {Schoop}}]{Klemenz2020}%
  \BibitemOpen
  \bibfield  {author} {\bibinfo {author} {\bibfnamefont {S.}~\bibnamefont {Klemenz}}, \bibinfo {author} {\bibfnamefont {A.~K.}\ \bibnamefont {Hay}}, \bibinfo {author} {\bibfnamefont {S.~M.~L.}\ \bibnamefont {Teicher}}, \bibinfo {author} {\bibfnamefont {A.}~\bibnamefont {Topp}}, \bibinfo {author} {\bibfnamefont {J.}~\bibnamefont {Cano}},\ and\ \bibinfo {author} {\bibfnamefont {L.~M.}\ \bibnamefont {Schoop}},\ }\href {https://doi.org/10.1021/jacs.0c01227} {\bibfield  {journal} {\bibinfo  {journal} {J. Am. Chem. Soc.}\ }\textbf {\bibinfo {volume} {142}},\ \bibinfo {pages} {6350} (\bibinfo {year} {2020})}\BibitemShut {NoStop}%
\bibitem [{\citenamefont {Wang}\ and\ \citenamefont {Hughbanks}(1995)}]{wang1995}%
  \BibitemOpen
  \bibfield  {author} {\bibinfo {author} {\bibfnamefont {C.}~\bibnamefont {Wang}}\ and\ \bibinfo {author} {\bibfnamefont {T.}~\bibnamefont {Hughbanks}},\ }\href {https://doi.org/10.1021/ic00126a024} {\bibfield  {journal} {\bibinfo  {journal} {Inorg. Chem.}\ }\textbf {\bibinfo {volume} {34}},\ \bibinfo {pages} {5524} (\bibinfo {year} {1995})}\BibitemShut {NoStop}%
\bibitem [{\citenamefont {Xu}\ \emph {et~al.}(2015)\citenamefont {Xu}, \citenamefont {Song}, \citenamefont {Nie}, \citenamefont {Weng}, \citenamefont {Fang},\ and\ \citenamefont {Dai}}]{Xu2015}%
  \BibitemOpen
  \bibfield  {author} {\bibinfo {author} {\bibfnamefont {Q.}~\bibnamefont {Xu}}, \bibinfo {author} {\bibfnamefont {Z.}~\bibnamefont {Song}}, \bibinfo {author} {\bibfnamefont {S.}~\bibnamefont {Nie}}, \bibinfo {author} {\bibfnamefont {H.}~\bibnamefont {Weng}}, \bibinfo {author} {\bibfnamefont {Z.}~\bibnamefont {Fang}},\ and\ \bibinfo {author} {\bibfnamefont {X.}~\bibnamefont {Dai}},\ }\href {https://doi.org/10.1103/PhysRevB.92.205310} {\bibfield  {journal} {\bibinfo  {journal} {Phys. Rev. B}\ }\textbf {\bibinfo {volume} {92}},\ \bibinfo {pages} {205310} (\bibinfo {year} {2015})}\BibitemShut {NoStop}%
\bibitem [{\citenamefont {Lv}\ \emph {et~al.}(2016)\citenamefont {Lv}, \citenamefont {Zhang}, \citenamefont {Li}, \citenamefont {Yao}, \citenamefont {Chen}, \citenamefont {Zhou}, \citenamefont {Zhang}, \citenamefont {Lu},\ and\ \citenamefont {Chen}}]{lv2016}%
  \BibitemOpen
  \bibfield  {author} {\bibinfo {author} {\bibfnamefont {Y.-Y.}\ \bibnamefont {Lv}}, \bibinfo {author} {\bibfnamefont {B.-B.}\ \bibnamefont {Zhang}}, \bibinfo {author} {\bibfnamefont {X.}~\bibnamefont {Li}}, \bibinfo {author} {\bibfnamefont {S.-H.}\ \bibnamefont {Yao}}, \bibinfo {author} {\bibfnamefont {Y.}~\bibnamefont {Chen}}, \bibinfo {author} {\bibfnamefont {J.}~\bibnamefont {Zhou}}, \bibinfo {author} {\bibfnamefont {S.-T.}\ \bibnamefont {Zhang}}, \bibinfo {author} {\bibfnamefont {M.-H.}\ \bibnamefont {Lu}},\ and\ \bibinfo {author} {\bibfnamefont {Y.-F.}\ \bibnamefont {Chen}},\ }\href {https://pubs.aip.org/aip/apl/article/108/24/244101/971675/Extremely-large-and-significantly-anisotropic} {\bibfield  {journal} {\bibinfo  {journal} {Appl. Phys. Lett.}\ }\textbf {\bibinfo {volume} {108}} (\bibinfo {year} {2016})}\BibitemShut {NoStop}%
\bibitem [{\citenamefont {Yang}\ \emph {et~al.}(2021)\citenamefont {Yang}, \citenamefont {Song}, \citenamefont {Guo}, \citenamefont {Gao}, \citenamefont {Dong}, \citenamefont {Yu}, \citenamefont {Zheng}, \citenamefont {Kang},\ and\ \citenamefont {Zhang}}]{Yang2021}%
  \BibitemOpen
  \bibfield  {author} {\bibinfo {author} {\bibfnamefont {J.}~\bibnamefont {Yang}}, \bibinfo {author} {\bibfnamefont {Z.-Y.}\ \bibnamefont {Song}}, \bibinfo {author} {\bibfnamefont {L.}~\bibnamefont {Guo}}, \bibinfo {author} {\bibfnamefont {H.}~\bibnamefont {Gao}}, \bibinfo {author} {\bibfnamefont {Z.}~\bibnamefont {Dong}}, \bibinfo {author} {\bibfnamefont {Q.}~\bibnamefont {Yu}}, \bibinfo {author} {\bibfnamefont {R.-K.}\ \bibnamefont {Zheng}}, \bibinfo {author} {\bibfnamefont {T.-T.}\ \bibnamefont {Kang}},\ and\ \bibinfo {author} {\bibfnamefont {K.}~\bibnamefont {Zhang}},\ }\href {https://doi.org/https://doi.org/10.1021/acs.nanolett.1c01647} {\bibfield  {journal} {\bibinfo  {journal} {Nano Lett.}\ }\textbf {\bibinfo {volume} {21}},\ \bibinfo {pages} {10139} (\bibinfo {year} {2021})}\BibitemShut {NoStop}%
\bibitem [{\citenamefont {Singha}\ \emph {et~al.}(2017{\natexlab{a}})\citenamefont {Singha}, \citenamefont {Pariari}, \citenamefont {Satpati},\ and\ \citenamefont {Mandal}}]{Singha2017}%
  \BibitemOpen
  \bibfield  {author} {\bibinfo {author} {\bibfnamefont {R.}~\bibnamefont {Singha}}, \bibinfo {author} {\bibfnamefont {A.~K.}\ \bibnamefont {Pariari}}, \bibinfo {author} {\bibfnamefont {B.}~\bibnamefont {Satpati}},\ and\ \bibinfo {author} {\bibfnamefont {P.}~\bibnamefont {Mandal}},\ }\href {https://doi.org/10.1073/pnas.1618004114} {\bibfield  {journal} {\bibinfo  {journal} {Proc. Natl. Acad. Sci.}\ }\textbf {\bibinfo {volume} {114}},\ \bibinfo {pages} {2468} (\bibinfo {year} {2017}{\natexlab{a}})}\BibitemShut {NoStop}%
\bibitem [{\citenamefont {Hu}\ \emph {et~al.}(2017)\citenamefont {Hu}, \citenamefont {Zhu}, \citenamefont {Graf}, \citenamefont {Tang}, \citenamefont {Liu},\ and\ \citenamefont {Mao}}]{Hu2017}%
  \BibitemOpen
  \bibfield  {author} {\bibinfo {author} {\bibfnamefont {J.}~\bibnamefont {Hu}}, \bibinfo {author} {\bibfnamefont {Y.~L.}\ \bibnamefont {Zhu}}, \bibinfo {author} {\bibfnamefont {D.}~\bibnamefont {Graf}}, \bibinfo {author} {\bibfnamefont {Z.~J.}\ \bibnamefont {Tang}}, \bibinfo {author} {\bibfnamefont {J.~Y.}\ \bibnamefont {Liu}},\ and\ \bibinfo {author} {\bibfnamefont {Z.~Q.}\ \bibnamefont {Mao}},\ }\href {https://doi.org/10.1103/PhysRevB.95.205134} {\bibfield  {journal} {\bibinfo  {journal} {Phys. Rev. B}\ }\textbf {\bibinfo {volume} {95}},\ \bibinfo {pages} {205134} (\bibinfo {year} {2017})}\BibitemShut {NoStop}%
\bibitem [{\citenamefont {Wang}\ \emph {et~al.}(2021)\citenamefont {Wang}, \citenamefont {Qian}, \citenamefont {Yang}, \citenamefont {Chen}, \citenamefont {Li}, \citenamefont {Tan}, \citenamefont {Cai}, \citenamefont {Zhao}, \citenamefont {Gao}, \citenamefont {Feng}, \citenamefont {Kumar}, \citenamefont {Schwier}, \citenamefont {Zhao}, \citenamefont {Weng}, \citenamefont {Shi}, \citenamefont {Wang}, \citenamefont {Song}, \citenamefont {Huang}, \citenamefont {Shimada}, \citenamefont {Xu}, \citenamefont {Zhou},\ and\ \citenamefont {Liu}}]{Wang2021}%
  \BibitemOpen
  \bibfield  {author} {\bibinfo {author} {\bibfnamefont {Y.}~\bibnamefont {Wang}}, \bibinfo {author} {\bibfnamefont {Y.}~\bibnamefont {Qian}}, \bibinfo {author} {\bibfnamefont {M.}~\bibnamefont {Yang}}, \bibinfo {author} {\bibfnamefont {H.}~\bibnamefont {Chen}}, \bibinfo {author} {\bibfnamefont {C.}~\bibnamefont {Li}}, \bibinfo {author} {\bibfnamefont {Z.}~\bibnamefont {Tan}}, \bibinfo {author} {\bibfnamefont {Y.}~\bibnamefont {Cai}}, \bibinfo {author} {\bibfnamefont {W.}~\bibnamefont {Zhao}}, \bibinfo {author} {\bibfnamefont {S.}~\bibnamefont {Gao}}, \bibinfo {author} {\bibfnamefont {Y.}~\bibnamefont {Feng}}, \bibinfo {author} {\bibfnamefont {S.}~\bibnamefont {Kumar}}, \bibinfo {author} {\bibfnamefont {E.~F.}\ \bibnamefont {Schwier}}, \bibinfo {author} {\bibfnamefont {L.}~\bibnamefont {Zhao}}, \bibinfo {author} {\bibfnamefont {H.}~\bibnamefont {Weng}}, \bibinfo {author} {\bibfnamefont {Y.}~\bibnamefont {Shi}}, \bibinfo {author} {\bibfnamefont {G.}~\bibnamefont {Wang}}, \bibinfo {author} {\bibfnamefont
  {Y.}~\bibnamefont {Song}}, \bibinfo {author} {\bibfnamefont {Y.}~\bibnamefont {Huang}}, \bibinfo {author} {\bibfnamefont {K.}~\bibnamefont {Shimada}}, \bibinfo {author} {\bibfnamefont {Z.}~\bibnamefont {Xu}}, \bibinfo {author} {\bibfnamefont {X.~J.}\ \bibnamefont {Zhou}},\ and\ \bibinfo {author} {\bibfnamefont {G.}~\bibnamefont {Liu}},\ }\href {https://doi.org/10.1103/PhysRevB.103.125131} {\bibfield  {journal} {\bibinfo  {journal} {Phys. Rev. B}\ }\textbf {\bibinfo {volume} {103}},\ \bibinfo {pages} {125131} (\bibinfo {year} {2021})}\BibitemShut {NoStop}%
\bibitem [{\citenamefont {Lv}\ \emph {et~al.}(2019)\citenamefont {Lv}, \citenamefont {Chen}, \citenamefont {Qiao}, \citenamefont {Ma}, \citenamefont {Yang}, \citenamefont {Li}, \citenamefont {Wang}, \citenamefont {Tao},\ and\ \citenamefont {Xu}}]{lv2019}%
  \BibitemOpen
  \bibfield  {author} {\bibinfo {author} {\bibfnamefont {B.}~\bibnamefont {Lv}}, \bibinfo {author} {\bibfnamefont {J.}~\bibnamefont {Chen}}, \bibinfo {author} {\bibfnamefont {L.}~\bibnamefont {Qiao}}, \bibinfo {author} {\bibfnamefont {J.}~\bibnamefont {Ma}}, \bibinfo {author} {\bibfnamefont {X.}~\bibnamefont {Yang}}, \bibinfo {author} {\bibfnamefont {M.}~\bibnamefont {Li}}, \bibinfo {author} {\bibfnamefont {M.}~\bibnamefont {Wang}}, \bibinfo {author} {\bibfnamefont {Q.}~\bibnamefont {Tao}},\ and\ \bibinfo {author} {\bibfnamefont {Z.-A.}\ \bibnamefont {Xu}},\ }\href {https://doi.org/10.1088/1361-648X/ab2498} {\bibfield  {journal} {\bibinfo  {journal} {J. Phys.: Condens. Matter}\ }\textbf {\bibinfo {volume} {31}},\ \bibinfo {pages} {355601} (\bibinfo {year} {2019})}\BibitemShut {NoStop}%
\bibitem [{\citenamefont {Hosen}\ \emph {et~al.}(2018)\citenamefont {Hosen}, \citenamefont {Dhakal}, \citenamefont {Dimitri}, \citenamefont {Maldonado}, \citenamefont {Aperis}, \citenamefont {Kabir}, \citenamefont {Sims}, \citenamefont {Riseborough}, \citenamefont {Oppeneer}, \citenamefont {Kaczorowski}, \citenamefont {Durakiewicz},\ and\ \citenamefont {Neupane}}]{Hosen2018}%
  \BibitemOpen
  \bibfield  {author} {\bibinfo {author} {\bibfnamefont {M.~M.}\ \bibnamefont {Hosen}}, \bibinfo {author} {\bibfnamefont {G.}~\bibnamefont {Dhakal}}, \bibinfo {author} {\bibfnamefont {K.}~\bibnamefont {Dimitri}}, \bibinfo {author} {\bibfnamefont {P.}~\bibnamefont {Maldonado}}, \bibinfo {author} {\bibfnamefont {A.}~\bibnamefont {Aperis}}, \bibinfo {author} {\bibfnamefont {F.}~\bibnamefont {Kabir}}, \bibinfo {author} {\bibfnamefont {C.}~\bibnamefont {Sims}}, \bibinfo {author} {\bibfnamefont {P.}~\bibnamefont {Riseborough}}, \bibinfo {author} {\bibfnamefont {P.~M.}\ \bibnamefont {Oppeneer}}, \bibinfo {author} {\bibfnamefont {D.}~\bibnamefont {Kaczorowski}}, \bibinfo {author} {\bibfnamefont {T.}~\bibnamefont {Durakiewicz}},\ and\ \bibinfo {author} {\bibfnamefont {M.}~\bibnamefont {Neupane}},\ }\href {https://doi.org/10.1038/s41598-018-31296-7} {\bibfield  {journal} {\bibinfo  {journal} {Sci. Rep.}\ }\textbf {\bibinfo {volume} {8}},\ \bibinfo {pages} {13283} (\bibinfo {year} {2018})}\BibitemShut {NoStop}%
\bibitem [{\citenamefont {Regmi}\ \emph {et~al.}(2022)\citenamefont {Regmi}, \citenamefont {Dhakal}, \citenamefont {Kabeer}, \citenamefont {Harrison}, \citenamefont {Kabir}, \citenamefont {Sakhya}, \citenamefont {Gofryk}, \citenamefont {Kaczorowski}, \citenamefont {Oppeneer},\ and\ \citenamefont {Neupane}}]{Regmi2022}%
  \BibitemOpen
  \bibfield  {author} {\bibinfo {author} {\bibfnamefont {S.}~\bibnamefont {Regmi}}, \bibinfo {author} {\bibfnamefont {G.}~\bibnamefont {Dhakal}}, \bibinfo {author} {\bibfnamefont {F.~C.}\ \bibnamefont {Kabeer}}, \bibinfo {author} {\bibfnamefont {N.}~\bibnamefont {Harrison}}, \bibinfo {author} {\bibfnamefont {F.}~\bibnamefont {Kabir}}, \bibinfo {author} {\bibfnamefont {A.~P.}\ \bibnamefont {Sakhya}}, \bibinfo {author} {\bibfnamefont {K.}~\bibnamefont {Gofryk}}, \bibinfo {author} {\bibfnamefont {D.}~\bibnamefont {Kaczorowski}}, \bibinfo {author} {\bibfnamefont {P.~M.}\ \bibnamefont {Oppeneer}},\ and\ \bibinfo {author} {\bibfnamefont {M.}~\bibnamefont {Neupane}},\ }\href {https://doi.org/10.1103/PhysRevMaterials.6.L031201} {\bibfield  {journal} {\bibinfo  {journal} {Phys. Rev. Mater.}\ }\textbf {\bibinfo {volume} {6}},\ \bibinfo {pages} {L031201} (\bibinfo {year} {2022})}\BibitemShut {NoStop}%
\bibitem [{\citenamefont {Chen}\ \emph {et~al.}(2017)\citenamefont {Chen}, \citenamefont {Lai}, \citenamefont {Chiu}, \citenamefont {Steven}, \citenamefont {Besara}, \citenamefont {Graf}, \citenamefont {Siegrist}, \citenamefont {Albrecht-Schmitt}, \citenamefont {Balicas},\ and\ \citenamefont {Baumbach}}]{Chen2017}%
  \BibitemOpen
  \bibfield  {author} {\bibinfo {author} {\bibfnamefont {K.-W.}\ \bibnamefont {Chen}}, \bibinfo {author} {\bibfnamefont {Y.}~\bibnamefont {Lai}}, \bibinfo {author} {\bibfnamefont {Y.-C.}\ \bibnamefont {Chiu}}, \bibinfo {author} {\bibfnamefont {S.}~\bibnamefont {Steven}}, \bibinfo {author} {\bibfnamefont {T.}~\bibnamefont {Besara}}, \bibinfo {author} {\bibfnamefont {D.}~\bibnamefont {Graf}}, \bibinfo {author} {\bibfnamefont {T.}~\bibnamefont {Siegrist}}, \bibinfo {author} {\bibfnamefont {T.~E.}\ \bibnamefont {Albrecht-Schmitt}}, \bibinfo {author} {\bibfnamefont {L.}~\bibnamefont {Balicas}},\ and\ \bibinfo {author} {\bibfnamefont {R.~E.}\ \bibnamefont {Baumbach}},\ }\href {https://doi.org/10.1103/PhysRevB.96.014421} {\bibfield  {journal} {\bibinfo  {journal} {Phys. Rev. B}\ }\textbf {\bibinfo {volume} {96}},\ \bibinfo {pages} {014421} (\bibinfo {year} {2017})}\BibitemShut {NoStop}%
\bibitem [{\citenamefont {Pandey}\ \emph {et~al.}(2020)\citenamefont {Pandey}, \citenamefont {Basnet}, \citenamefont {Wegner}, \citenamefont {Acharya}, \citenamefont {Nabi}, \citenamefont {Liu}, \citenamefont {Wang}, \citenamefont {Takahashi}, \citenamefont {Da},\ and\ \citenamefont {Hu}}]{Pandey2020}%
  \BibitemOpen
  \bibfield  {author} {\bibinfo {author} {\bibfnamefont {K.}~\bibnamefont {Pandey}}, \bibinfo {author} {\bibfnamefont {R.}~\bibnamefont {Basnet}}, \bibinfo {author} {\bibfnamefont {A.}~\bibnamefont {Wegner}}, \bibinfo {author} {\bibfnamefont {G.}~\bibnamefont {Acharya}}, \bibinfo {author} {\bibfnamefont {M.~R.~U.}\ \bibnamefont {Nabi}}, \bibinfo {author} {\bibfnamefont {J.}~\bibnamefont {Liu}}, \bibinfo {author} {\bibfnamefont {J.}~\bibnamefont {Wang}}, \bibinfo {author} {\bibfnamefont {Y.~K.}\ \bibnamefont {Takahashi}}, \bibinfo {author} {\bibfnamefont {B.}~\bibnamefont {Da}},\ and\ \bibinfo {author} {\bibfnamefont {J.}~\bibnamefont {Hu}},\ }\href {https://doi.org/10.1103/PhysRevB.101.235161} {\bibfield  {journal} {\bibinfo  {journal} {Phys. Rev. B}\ }\textbf {\bibinfo {volume} {101}},\ \bibinfo {pages} {235161} (\bibinfo {year} {2020})}\BibitemShut {NoStop}%
\bibitem [{\citenamefont {Lei}\ \emph {et~al.}(2019)\citenamefont {Lei}, \citenamefont {Duppel}, \citenamefont {Lippmann}, \citenamefont {Nuss}, \citenamefont {Lotsch},\ and\ \citenamefont {Schoop}}]{Lei2019}%
  \BibitemOpen
  \bibfield  {author} {\bibinfo {author} {\bibfnamefont {S.}~\bibnamefont {Lei}}, \bibinfo {author} {\bibfnamefont {V.}~\bibnamefont {Duppel}}, \bibinfo {author} {\bibfnamefont {J.~M.}\ \bibnamefont {Lippmann}}, \bibinfo {author} {\bibfnamefont {J.}~\bibnamefont {Nuss}}, \bibinfo {author} {\bibfnamefont {B.~V.}\ \bibnamefont {Lotsch}},\ and\ \bibinfo {author} {\bibfnamefont {L.~M.}\ \bibnamefont {Schoop}},\ }\href {https://doi.org/https://doi.org/10.1002/qute.201900045} {\bibfield  {journal} {\bibinfo  {journal} {Adv. Quantum Technol.}\ }\textbf {\bibinfo {volume} {2}},\ \bibinfo {pages} {1900045} (\bibinfo {year} {2019})}\BibitemShut {NoStop}%
\bibitem [{\citenamefont {Lei}\ \emph {et~al.}(2021)\citenamefont {Lei}, \citenamefont {Teicher}, \citenamefont {Topp}, \citenamefont {Cai}, \citenamefont {Lin}, \citenamefont {Cheng}, \citenamefont {Salters}, \citenamefont {Rodolakis}, \citenamefont {McChesney}, \citenamefont {Lapidus}, \citenamefont {Yao}, \citenamefont {Krivenkov}, \citenamefont {Marchenko}, \citenamefont {Varykhalov}, \citenamefont {Ast}, \citenamefont {Car}, \citenamefont {Cano}, \citenamefont {Vergniory}, \citenamefont {Ong},\ and\ \citenamefont {Schoop}}]{Lei2021}%
  \BibitemOpen
  \bibfield  {author} {\bibinfo {author} {\bibfnamefont {S.}~\bibnamefont {Lei}}, \bibinfo {author} {\bibfnamefont {S.~M.~L.}\ \bibnamefont {Teicher}}, \bibinfo {author} {\bibfnamefont {A.}~\bibnamefont {Topp}}, \bibinfo {author} {\bibfnamefont {K.}~\bibnamefont {Cai}}, \bibinfo {author} {\bibfnamefont {J.}~\bibnamefont {Lin}}, \bibinfo {author} {\bibfnamefont {G.}~\bibnamefont {Cheng}}, \bibinfo {author} {\bibfnamefont {T.~H.}\ \bibnamefont {Salters}}, \bibinfo {author} {\bibfnamefont {F.}~\bibnamefont {Rodolakis}}, \bibinfo {author} {\bibfnamefont {J.~L.}\ \bibnamefont {McChesney}}, \bibinfo {author} {\bibfnamefont {S.}~\bibnamefont {Lapidus}}, \bibinfo {author} {\bibfnamefont {N.}~\bibnamefont {Yao}}, \bibinfo {author} {\bibfnamefont {M.}~\bibnamefont {Krivenkov}}, \bibinfo {author} {\bibfnamefont {D.}~\bibnamefont {Marchenko}}, \bibinfo {author} {\bibfnamefont {A.}~\bibnamefont {Varykhalov}}, \bibinfo {author} {\bibfnamefont {C.~R.}\ \bibnamefont {Ast}}, \bibinfo {author} {\bibfnamefont {R.}~\bibnamefont
  {Car}}, \bibinfo {author} {\bibfnamefont {J.}~\bibnamefont {Cano}}, \bibinfo {author} {\bibfnamefont {M.~G.}\ \bibnamefont {Vergniory}}, \bibinfo {author} {\bibfnamefont {N.~P.}\ \bibnamefont {Ong}},\ and\ \bibinfo {author} {\bibfnamefont {L.~M.}\ \bibnamefont {Schoop}},\ }\href {https://doi.org/https://doi.org/10.1002/adma.202101591} {\bibfield  {journal} {\bibinfo  {journal} {Adv. Mater.}\ }\textbf {\bibinfo {volume} {33}},\ \bibinfo {pages} {2101591} (\bibinfo {year} {2021})}\BibitemShut {NoStop}%
\bibitem [{\citenamefont {Hosen}\ \emph {et~al.}(2017)\citenamefont {Hosen}, \citenamefont {Dimitri}, \citenamefont {Belopolski}, \citenamefont {Maldonado}, \citenamefont {Sankar}, \citenamefont {Dhakal}, \citenamefont {Dhakal}, \citenamefont {Cole}, \citenamefont {Oppeneer}, \citenamefont {Kaczorowski}, \citenamefont {Chou}, \citenamefont {Hasan}, \citenamefont {Durakiewicz},\ and\ \citenamefont {Neupane}}]{Hosen2017}%
  \BibitemOpen
  \bibfield  {author} {\bibinfo {author} {\bibfnamefont {M.~M.}\ \bibnamefont {Hosen}}, \bibinfo {author} {\bibfnamefont {K.}~\bibnamefont {Dimitri}}, \bibinfo {author} {\bibfnamefont {I.}~\bibnamefont {Belopolski}}, \bibinfo {author} {\bibfnamefont {P.}~\bibnamefont {Maldonado}}, \bibinfo {author} {\bibfnamefont {R.}~\bibnamefont {Sankar}}, \bibinfo {author} {\bibfnamefont {N.}~\bibnamefont {Dhakal}}, \bibinfo {author} {\bibfnamefont {G.}~\bibnamefont {Dhakal}}, \bibinfo {author} {\bibfnamefont {T.}~\bibnamefont {Cole}}, \bibinfo {author} {\bibfnamefont {P.~M.}\ \bibnamefont {Oppeneer}}, \bibinfo {author} {\bibfnamefont {D.}~\bibnamefont {Kaczorowski}}, \bibinfo {author} {\bibfnamefont {F.}~\bibnamefont {Chou}}, \bibinfo {author} {\bibfnamefont {M.~Z.}\ \bibnamefont {Hasan}}, \bibinfo {author} {\bibfnamefont {T.}~\bibnamefont {Durakiewicz}},\ and\ \bibinfo {author} {\bibfnamefont {M.}~\bibnamefont {Neupane}},\ }\href {https://doi.org/10.1103/PhysRevB.95.161101} {\bibfield  {journal} {\bibinfo  {journal}
  {Phys. Rev. B}\ }\textbf {\bibinfo {volume} {95}},\ \bibinfo {pages} {161101} (\bibinfo {year} {2017})}\BibitemShut {NoStop}%
\bibitem [{\citenamefont {Song}\ \emph {et~al.}(2021)\citenamefont {Song}, \citenamefont {Song}, \citenamefont {Li}, \citenamefont {Wang}, \citenamefont {Wang}, \citenamefont {Zhang}, \citenamefont {Han}, \citenamefont {Cao}, \citenamefont {Xiong},\ and\ \citenamefont {Liu}}]{Song2021}%
  \BibitemOpen
  \bibfield  {author} {\bibinfo {author} {\bibfnamefont {J.}~\bibnamefont {Song}}, \bibinfo {author} {\bibfnamefont {M.}~\bibnamefont {Song}}, \bibinfo {author} {\bibfnamefont {Z.}~\bibnamefont {Li}}, \bibinfo {author} {\bibfnamefont {J.}~\bibnamefont {Wang}}, \bibinfo {author} {\bibfnamefont {Y.}~\bibnamefont {Wang}}, \bibinfo {author} {\bibfnamefont {L.}~\bibnamefont {Zhang}}, \bibinfo {author} {\bibfnamefont {Y.}~\bibnamefont {Han}}, \bibinfo {author} {\bibfnamefont {L.}~\bibnamefont {Cao}}, \bibinfo {author} {\bibfnamefont {Y.}~\bibnamefont {Xiong}},\ and\ \bibinfo {author} {\bibfnamefont {D.}~\bibnamefont {Liu}},\ }\href {https://doi.org/10.1103/PhysRevB.103.165141} {\bibfield  {journal} {\bibinfo  {journal} {Phys. Rev. B}\ }\textbf {\bibinfo {volume} {103}},\ \bibinfo {pages} {165141} (\bibinfo {year} {2021})}\BibitemShut {NoStop}%
\bibitem [{\citenamefont {Sankar}\ \emph {et~al.}(2019)\citenamefont {Sankar}, \citenamefont {Muthuselvam}, \citenamefont {Babu}, \citenamefont {Murugan}, \citenamefont {Rajagopal}, \citenamefont {Kumar}, \citenamefont {Wu}, \citenamefont {Wen}, \citenamefont {Lee}, \citenamefont {Guo},\ and\ \citenamefont {Chou}}]{Sankar2019}%
  \BibitemOpen
  \bibfield  {author} {\bibinfo {author} {\bibfnamefont {R.}~\bibnamefont {Sankar}}, \bibinfo {author} {\bibfnamefont {I.~P.}\ \bibnamefont {Muthuselvam}}, \bibinfo {author} {\bibfnamefont {K.~R.}\ \bibnamefont {Babu}}, \bibinfo {author} {\bibfnamefont {G.~S.}\ \bibnamefont {Murugan}}, \bibinfo {author} {\bibfnamefont {K.}~\bibnamefont {Rajagopal}}, \bibinfo {author} {\bibfnamefont {R.}~\bibnamefont {Kumar}}, \bibinfo {author} {\bibfnamefont {T.-C.}\ \bibnamefont {Wu}}, \bibinfo {author} {\bibfnamefont {C.-Y.}\ \bibnamefont {Wen}}, \bibinfo {author} {\bibfnamefont {W.-L.}\ \bibnamefont {Lee}}, \bibinfo {author} {\bibfnamefont {G.-Y.}\ \bibnamefont {Guo}},\ and\ \bibinfo {author} {\bibfnamefont {F.-C.}\ \bibnamefont {Chou}},\ }\href {https://doi.org/10.1021/acs.inorgchem.9b01698} {\bibfield  {journal} {\bibinfo  {journal} {Inorg. Chem.}\ }\textbf {\bibinfo {volume} {58}},\ \bibinfo {pages} {11730} (\bibinfo {year} {2019})}\BibitemShut {NoStop}%
\bibitem [{\citenamefont {Pandey}\ \emph {et~al.}(2022)\citenamefont {Pandey}, \citenamefont {Sayler}, \citenamefont {Basnet}, \citenamefont {Sakon}, \citenamefont {Wang},\ and\ \citenamefont {Hu}}]{Pandey2022}%
  \BibitemOpen
  \bibfield  {author} {\bibinfo {author} {\bibfnamefont {K.}~\bibnamefont {Pandey}}, \bibinfo {author} {\bibfnamefont {L.}~\bibnamefont {Sayler}}, \bibinfo {author} {\bibfnamefont {R.}~\bibnamefont {Basnet}}, \bibinfo {author} {\bibfnamefont {J.}~\bibnamefont {Sakon}}, \bibinfo {author} {\bibfnamefont {F.}~\bibnamefont {Wang}},\ and\ \bibinfo {author} {\bibfnamefont {J.}~\bibnamefont {Hu}},\ }\href {https://doi.org/https://doi.org/10.3390/cryst12111663} {\bibfield  {journal} {\bibinfo  {journal} {Cryst.}\ }\textbf {\bibinfo {volume} {12}},\ \bibinfo {pages} {1663} (\bibinfo {year} {2022})}\BibitemShut {NoStop}%
\bibitem [{\citenamefont {{Bruker AXS, Germany}}(2009)}]{topas}%
  \BibitemOpen
  \bibfield  {author} {\bibinfo {author} {\bibnamefont {{Bruker AXS, Germany}}},\ }\href {https://www.bruker.com/products/x-ray-diffraction-and-elemental-analysis/x-ray-diffraction/xrd-software/topas.html} {\bibinfo {title} {Topas 4.2}} (\bibinfo {year} {2009})\BibitemShut {NoStop}%
\bibitem [{\citenamefont {Gebauer}\ \emph {et~al.}(2021)\citenamefont {Gebauer}, \citenamefont {Poddig}, \citenamefont {Corredor-Bohorquez}, \citenamefont {Menshchikova}, \citenamefont {Rusinov}, \citenamefont {Golub}, \citenamefont {Caglieris}, \citenamefont {Benndorf}, \citenamefont {Lindemann}, \citenamefont {Chulkov}, \citenamefont {Wolter}, \citenamefont {Bernd~Büchner}, \citenamefont {Doert},\ and\ \citenamefont {Isaeva}}]{Gebauer2021}%
  \BibitemOpen
  \bibfield  {author} {\bibinfo {author} {\bibfnamefont {P.}~\bibnamefont {Gebauer}}, \bibinfo {author} {\bibfnamefont {H.}~\bibnamefont {Poddig}}, \bibinfo {author} {\bibfnamefont {L.~T.}\ \bibnamefont {Corredor-Bohorquez}}, \bibinfo {author} {\bibfnamefont {T.~V.}\ \bibnamefont {Menshchikova}}, \bibinfo {author} {\bibfnamefont {I.~P.}\ \bibnamefont {Rusinov}}, \bibinfo {author} {\bibfnamefont {P.}~\bibnamefont {Golub}}, \bibinfo {author} {\bibfnamefont {F.}~\bibnamefont {Caglieris}}, \bibinfo {author} {\bibfnamefont {C.}~\bibnamefont {Benndorf}}, \bibinfo {author} {\bibfnamefont {T.}~\bibnamefont {Lindemann}}, \bibinfo {author} {\bibfnamefont {E.~V.}\ \bibnamefont {Chulkov}}, \bibinfo {author} {\bibfnamefont {A.~U.~B.}\ \bibnamefont {Wolter}}, \bibinfo {author} {\bibfnamefont {B.}~\bibnamefont {Bernd~Büchner}}, \bibinfo {author} {\bibfnamefont {T.}~\bibnamefont {Doert}},\ and\ \bibinfo {author} {\bibfnamefont {A.}~\bibnamefont {Isaeva}},\ }\href {https://doi.org/10.1021/acs.chemmater.0c04649} {\bibfield
  {journal} {\bibinfo  {journal} {Chem. Mater.}\ }\textbf {\bibinfo {volume} {33}},\ \bibinfo {pages} {2420} (\bibinfo {year} {2021})}\BibitemShut {NoStop}%
\bibitem [{\citenamefont {Ram}\ \emph {et~al.}(2023)\citenamefont {Ram}, \citenamefont {Singh}, \citenamefont {Hooda}, \citenamefont {Singh}, \citenamefont {Kanchana}, \citenamefont {Kaczorowski},\ and\ \citenamefont {Hossain}}]{Ram2023}%
  \BibitemOpen
  \bibfield  {author} {\bibinfo {author} {\bibfnamefont {D.}~\bibnamefont {Ram}}, \bibinfo {author} {\bibfnamefont {J.}~\bibnamefont {Singh}}, \bibinfo {author} {\bibfnamefont {M.~K.}\ \bibnamefont {Hooda}}, \bibinfo {author} {\bibfnamefont {K.}~\bibnamefont {Singh}}, \bibinfo {author} {\bibfnamefont {V.}~\bibnamefont {Kanchana}}, \bibinfo {author} {\bibfnamefont {D.}~\bibnamefont {Kaczorowski}},\ and\ \bibinfo {author} {\bibfnamefont {Z.}~\bibnamefont {Hossain}},\ }\href {https://doi.org/10.1103/PhysRevB.108.235107} {\bibfield  {journal} {\bibinfo  {journal} {Phys. Rev. B}\ }\textbf {\bibinfo {volume} {108}},\ \bibinfo {pages} {235107} (\bibinfo {year} {2023})}\BibitemShut {NoStop}%
\bibitem [{\citenamefont {Sahu}\ \emph {et~al.}(2021)\citenamefont {Sahu}, \citenamefont {Fobasso},\ and\ \citenamefont {Strydom}}]{Sahu2021}%
  \BibitemOpen
  \bibfield  {author} {\bibinfo {author} {\bibfnamefont {B.}~\bibnamefont {Sahu}}, \bibinfo {author} {\bibfnamefont {R.~D.}\ \bibnamefont {Fobasso}},\ and\ \bibinfo {author} {\bibfnamefont {A.~M.}\ \bibnamefont {Strydom}},\ }\href {https://doi.org/https://doi.org/10.1016/j.intermet.2021.107214} {\bibfield  {journal} {\bibinfo  {journal} {Intermetallics}\ }\textbf {\bibinfo {volume} {135}},\ \bibinfo {pages} {107214} (\bibinfo {year} {2021})}\BibitemShut {NoStop}%
\bibitem [{\citenamefont {Chakraborty}\ \emph {et~al.}(2022)\citenamefont {Chakraborty}, \citenamefont {Gupta}, \citenamefont {Pakhira}, \citenamefont {Choudhary}, \citenamefont {Biswas}, \citenamefont {Mudryk}, \citenamefont {Pecharsky}, \citenamefont {Johnson},\ and\ \citenamefont {Mazumdar}}]{Chakraborty2022}%
  \BibitemOpen
  \bibfield  {author} {\bibinfo {author} {\bibfnamefont {S.}~\bibnamefont {Chakraborty}}, \bibinfo {author} {\bibfnamefont {S.}~\bibnamefont {Gupta}}, \bibinfo {author} {\bibfnamefont {S.}~\bibnamefont {Pakhira}}, \bibinfo {author} {\bibfnamefont {R.}~\bibnamefont {Choudhary}}, \bibinfo {author} {\bibfnamefont {A.}~\bibnamefont {Biswas}}, \bibinfo {author} {\bibfnamefont {Y.}~\bibnamefont {Mudryk}}, \bibinfo {author} {\bibfnamefont {V.~K.}\ \bibnamefont {Pecharsky}}, \bibinfo {author} {\bibfnamefont {D.~D.}\ \bibnamefont {Johnson}},\ and\ \bibinfo {author} {\bibfnamefont {C.}~\bibnamefont {Mazumdar}},\ }\href {https://doi.org/10.1103/PhysRevB.106.224427} {\bibfield  {journal} {\bibinfo  {journal} {Phys. Rev. B}\ }\textbf {\bibinfo {volume} {106}},\ \bibinfo {pages} {224427} (\bibinfo {year} {2022})}\BibitemShut {NoStop}%
\bibitem [{\citenamefont {Muthuselvam}\ \emph {et~al.}(2019)\citenamefont {Muthuselvam}, \citenamefont {Nehru}, \citenamefont {Babu}, \citenamefont {Saranya}, \citenamefont {Kaul}, \citenamefont {Chen}, \citenamefont {Chen}, \citenamefont {Liu}, \citenamefont {Guo}, \citenamefont {Xiu},\ and\ \citenamefont {Sankar}}]{Muthuselvam2019}%
  \BibitemOpen
  \bibfield  {author} {\bibinfo {author} {\bibfnamefont {I.~P.}\ \bibnamefont {Muthuselvam}}, \bibinfo {author} {\bibfnamefont {R.}~\bibnamefont {Nehru}}, \bibinfo {author} {\bibfnamefont {K.~R.}\ \bibnamefont {Babu}}, \bibinfo {author} {\bibfnamefont {K.}~\bibnamefont {Saranya}}, \bibinfo {author} {\bibfnamefont {S.}~\bibnamefont {Kaul}}, \bibinfo {author} {\bibfnamefont {S.-M.}\ \bibnamefont {Chen}}, \bibinfo {author} {\bibfnamefont {W.-T.}\ \bibnamefont {Chen}}, \bibinfo {author} {\bibfnamefont {Y.}~\bibnamefont {Liu}}, \bibinfo {author} {\bibfnamefont {G.-Y.}\ \bibnamefont {Guo}}, \bibinfo {author} {\bibfnamefont {F.~X.}\ \bibnamefont {Xiu}},\ and\ \bibinfo {author} {\bibfnamefont {R.}~\bibnamefont {Sankar}},\ }\href {https://doi.org/10.1088/1361-648X/ab1570} {\bibfield  {journal} {\bibinfo  {journal} {J. Phys.: Condens. Matter}\ }\textbf {\bibinfo {volume} {31}},\ \bibinfo {pages} {285802} (\bibinfo {year} {2019})}\BibitemShut {NoStop}%
\bibitem [{\citenamefont {Li}\ \emph {et~al.}(2015)\citenamefont {Li}, \citenamefont {Wang}, \citenamefont {Cheng}, \citenamefont {Ren}, \citenamefont {Fang},\ and\ \citenamefont {Dou}}]{Li2015}%
  \BibitemOpen
  \bibfield  {author} {\bibinfo {author} {\bibfnamefont {G.}~\bibnamefont {Li}}, \bibinfo {author} {\bibfnamefont {J.}~\bibnamefont {Wang}}, \bibinfo {author} {\bibfnamefont {Z.}~\bibnamefont {Cheng}}, \bibinfo {author} {\bibfnamefont {Q.}~\bibnamefont {Ren}}, \bibinfo {author} {\bibfnamefont {C.}~\bibnamefont {Fang}},\ and\ \bibinfo {author} {\bibfnamefont {S.}~\bibnamefont {Dou}},\ }\href {https://pubs.aip.org/aip/apl/article/106/18/182405/27811/Large-entropy-change-accompanying-two-successive} {\bibfield  {journal} {\bibinfo  {journal} {Appl. Phys. Lett.}\ }\textbf {\bibinfo {volume} {106}} (\bibinfo {year} {2015})}\BibitemShut {NoStop}%
\bibitem [{\citenamefont {Banerjee}(1964)}]{Banerjee1964}%
  \BibitemOpen
  \bibfield  {author} {\bibinfo {author} {\bibfnamefont {B.}~\bibnamefont {Banerjee}},\ }\href {https://doi.org/https://doi.org/10.1016/0031-9163(64)91158-8} {\bibfield  {journal} {\bibinfo  {journal} {Phys. Lett.}\ }\textbf {\bibinfo {volume} {12}},\ \bibinfo {pages} {16} (\bibinfo {year} {1964})}\BibitemShut {NoStop}%
\bibitem [{\citenamefont {Mishra}\ and\ \citenamefont {Ganguli}(2024)}]{Mishra2024a}%
  \BibitemOpen
  \bibfield  {author} {\bibinfo {author} {\bibfnamefont {P.~K.}\ \bibnamefont {Mishra}}\ and\ \bibinfo {author} {\bibfnamefont {A.~K.}\ \bibnamefont {Ganguli}},\ }\href {https://doi.org/https://doi.org/10.1016/j.jssc.2024.124647} {\bibfield  {journal} {\bibinfo  {journal} {J. Solid State Chem.}\ }\textbf {\bibinfo {volume} {334}},\ \bibinfo {pages} {124647} (\bibinfo {year} {2024})}\BibitemShut {NoStop}%
\bibitem [{\citenamefont {Mishra}\ \emph {et~al.}(2024{\natexlab{a}})\citenamefont {Mishra}, \citenamefont {Singh}, \citenamefont {Gautam}, \citenamefont {Kumar}, \citenamefont {Umetsu},\ and\ \citenamefont {Ganguli}}]{Mishra2024c}%
  \BibitemOpen
  \bibfield  {author} {\bibinfo {author} {\bibfnamefont {P.~K.}\ \bibnamefont {Mishra}}, \bibinfo {author} {\bibfnamefont {H.}~\bibnamefont {Singh}}, \bibinfo {author} {\bibfnamefont {A.}~\bibnamefont {Gautam}}, \bibinfo {author} {\bibfnamefont {G.}~\bibnamefont {Kumar}}, \bibinfo {author} {\bibfnamefont {R.}~\bibnamefont {Umetsu}},\ and\ \bibinfo {author} {\bibfnamefont {A.~K.}\ \bibnamefont {Ganguli}},\ }\href {https://doi.org/10.1088/1402-4896/ad6e2d} {\bibfield  {journal} {\bibinfo  {journal} {Phys. Scr.}\ }\textbf {\bibinfo {volume} {99}},\ \bibinfo {pages} {095985} (\bibinfo {year} {2024}{\natexlab{a}})}\BibitemShut {NoStop}%
\bibitem [{\citenamefont {Biswas}\ \emph {et~al.}(2013)\citenamefont {Biswas}, \citenamefont {Chandra}, \citenamefont {Samanta}, \citenamefont {Phan}, \citenamefont {Das},\ and\ \citenamefont {Srikanth}}]{Biswas2013}%
  \BibitemOpen
  \bibfield  {author} {\bibinfo {author} {\bibfnamefont {A.}~\bibnamefont {Biswas}}, \bibinfo {author} {\bibfnamefont {S.}~\bibnamefont {Chandra}}, \bibinfo {author} {\bibfnamefont {T.}~\bibnamefont {Samanta}}, \bibinfo {author} {\bibfnamefont {M.}~\bibnamefont {Phan}}, \bibinfo {author} {\bibfnamefont {I.}~\bibnamefont {Das}},\ and\ \bibinfo {author} {\bibfnamefont {H.}~\bibnamefont {Srikanth}},\ }\href {https://pubs.aip.org/aip/jap/article/113/17/17A902/140385/The-universal-behavior-of-inverse-magnetocaloric} {\bibfield  {journal} {\bibinfo  {journal} {J. Appl. Phys.}\ }\textbf {\bibinfo {volume} {113}} (\bibinfo {year} {2013})}\BibitemShut {NoStop}%
\bibitem [{\citenamefont {Mishra}\ \emph {et~al.}(2024{\natexlab{b}})\citenamefont {Mishra}, \citenamefont {Gautam}, \citenamefont {Singh}, \citenamefont {Panda}, \citenamefont {Mohapatra},\ and\ \citenamefont {Ganguli}}]{Mishra2024b}%
  \BibitemOpen
  \bibfield  {author} {\bibinfo {author} {\bibfnamefont {P.~K.}\ \bibnamefont {Mishra}}, \bibinfo {author} {\bibfnamefont {A.}~\bibnamefont {Gautam}}, \bibinfo {author} {\bibfnamefont {H.}~\bibnamefont {Singh}}, \bibinfo {author} {\bibfnamefont {S.}~\bibnamefont {Panda}}, \bibinfo {author} {\bibfnamefont {N.}~\bibnamefont {Mohapatra}},\ and\ \bibinfo {author} {\bibfnamefont {A.~K.}\ \bibnamefont {Ganguli}},\ }\href {https://doi.org/10.1021/acs.chemmater.4c00453} {\bibfield  {journal} {\bibinfo  {journal} {Chem. Mater.}\ }\textbf {\bibinfo {volume} {36}},\ \bibinfo {pages} {5986} (\bibinfo {year} {2024}{\natexlab{b}})}\BibitemShut {NoStop}%
\bibitem [{\citenamefont {Kumar}\ and\ \citenamefont {Dhaka}(2020)}]{Kumar2020}%
  \BibitemOpen
  \bibfield  {author} {\bibinfo {author} {\bibfnamefont {A.}~\bibnamefont {Kumar}}\ and\ \bibinfo {author} {\bibfnamefont {R.~S.}\ \bibnamefont {Dhaka}},\ }\href {https://doi.org/10.1103/PhysRevB.101.094434} {\bibfield  {journal} {\bibinfo  {journal} {Phys. Rev. B}\ }\textbf {\bibinfo {volume} {101}},\ \bibinfo {pages} {094434} (\bibinfo {year} {2020})}\BibitemShut {NoStop}%
\bibitem [{\citenamefont {Schindler}\ \emph {et~al.}(2020)\citenamefont {Schindler}, \citenamefont {Galeski}, \citenamefont {Schnelle}, \citenamefont {Wawrzy\ifmmode~\acute{n}\else \'{n}\fi{}czak}, \citenamefont {Abdel-Haq}, \citenamefont {Guin}, \citenamefont {Kroder}, \citenamefont {Kumar}, \citenamefont {Fu}, \citenamefont {Borrmann}, \citenamefont {Shekhar}, \citenamefont {Felser}, \citenamefont {Meng}, \citenamefont {Grushin}, \citenamefont {Zhang}, \citenamefont {Sun},\ and\ \citenamefont {Gooth}}]{Schindler2020}%
  \BibitemOpen
  \bibfield  {author} {\bibinfo {author} {\bibfnamefont {C.}~\bibnamefont {Schindler}}, \bibinfo {author} {\bibfnamefont {S.}~\bibnamefont {Galeski}}, \bibinfo {author} {\bibfnamefont {W.}~\bibnamefont {Schnelle}}, \bibinfo {author} {\bibfnamefont {R.}~\bibnamefont {Wawrzy\ifmmode~\acute{n}\else \'{n}\fi{}czak}}, \bibinfo {author} {\bibfnamefont {W.}~\bibnamefont {Abdel-Haq}}, \bibinfo {author} {\bibfnamefont {S.~N.}\ \bibnamefont {Guin}}, \bibinfo {author} {\bibfnamefont {J.}~\bibnamefont {Kroder}}, \bibinfo {author} {\bibfnamefont {N.}~\bibnamefont {Kumar}}, \bibinfo {author} {\bibfnamefont {C.}~\bibnamefont {Fu}}, \bibinfo {author} {\bibfnamefont {H.}~\bibnamefont {Borrmann}}, \bibinfo {author} {\bibfnamefont {C.}~\bibnamefont {Shekhar}}, \bibinfo {author} {\bibfnamefont {C.}~\bibnamefont {Felser}}, \bibinfo {author} {\bibfnamefont {T.}~\bibnamefont {Meng}}, \bibinfo {author} {\bibfnamefont {A.~G.}\ \bibnamefont {Grushin}}, \bibinfo {author} {\bibfnamefont {Y.}~\bibnamefont {Zhang}}, \bibinfo {author}
  {\bibfnamefont {Y.}~\bibnamefont {Sun}},\ and\ \bibinfo {author} {\bibfnamefont {J.}~\bibnamefont {Gooth}},\ }\href {https://doi.org/10.1103/PhysRevB.101.125119} {\bibfield  {journal} {\bibinfo  {journal} {Phys. Rev. B}\ }\textbf {\bibinfo {volume} {101}},\ \bibinfo {pages} {125119} (\bibinfo {year} {2020})}\BibitemShut {NoStop}%
\bibitem [{\citenamefont {Singha}\ \emph {et~al.}(2017{\natexlab{b}})\citenamefont {Singha}, \citenamefont {Pariari}, \citenamefont {Satpati},\ and\ \citenamefont {Mandal}}]{Singha2017a}%
  \BibitemOpen
  \bibfield  {author} {\bibinfo {author} {\bibfnamefont {R.}~\bibnamefont {Singha}}, \bibinfo {author} {\bibfnamefont {A.}~\bibnamefont {Pariari}}, \bibinfo {author} {\bibfnamefont {B.}~\bibnamefont {Satpati}},\ and\ \bibinfo {author} {\bibfnamefont {P.}~\bibnamefont {Mandal}},\ }\href {https://doi.org/10.1103/PhysRevB.96.245138} {\bibfield  {journal} {\bibinfo  {journal} {Phys. Rev. B}\ }\textbf {\bibinfo {volume} {96}},\ \bibinfo {pages} {245138} (\bibinfo {year} {2017}{\natexlab{b}})}\BibitemShut {NoStop}%
\bibitem [{\citenamefont {Pandey}\ \emph {et~al.}(2021)\citenamefont {Pandey}, \citenamefont {Mondal}, \citenamefont {Villanova}, \citenamefont {Roll}, \citenamefont {Basnet}, \citenamefont {Wegner}, \citenamefont {Acharya}, \citenamefont {Nabi}, \citenamefont {Ghosh}, \citenamefont {Fujii}, \citenamefont {Wang}, \citenamefont {Da}, \citenamefont {Agarwal}, \citenamefont {Vobornik}, \citenamefont {Politano}, \citenamefont {Barraza-Lopez},\ and\ \citenamefont {Hu}}]{Pandey2021}%
  \BibitemOpen
  \bibfield  {author} {\bibinfo {author} {\bibfnamefont {K.}~\bibnamefont {Pandey}}, \bibinfo {author} {\bibfnamefont {D.}~\bibnamefont {Mondal}}, \bibinfo {author} {\bibfnamefont {J.~W.}\ \bibnamefont {Villanova}}, \bibinfo {author} {\bibfnamefont {J.}~\bibnamefont {Roll}}, \bibinfo {author} {\bibfnamefont {R.}~\bibnamefont {Basnet}}, \bibinfo {author} {\bibfnamefont {A.}~\bibnamefont {Wegner}}, \bibinfo {author} {\bibfnamefont {G.}~\bibnamefont {Acharya}}, \bibinfo {author} {\bibfnamefont {M.~R.~U.}\ \bibnamefont {Nabi}}, \bibinfo {author} {\bibfnamefont {B.}~\bibnamefont {Ghosh}}, \bibinfo {author} {\bibfnamefont {J.}~\bibnamefont {Fujii}}, \bibinfo {author} {\bibfnamefont {J.}~\bibnamefont {Wang}}, \bibinfo {author} {\bibfnamefont {B.}~\bibnamefont {Da}}, \bibinfo {author} {\bibfnamefont {A.}~\bibnamefont {Agarwal}}, \bibinfo {author} {\bibfnamefont {I.}~\bibnamefont {Vobornik}}, \bibinfo {author} {\bibfnamefont {A.}~\bibnamefont {Politano}}, \bibinfo {author} {\bibfnamefont {S.}~\bibnamefont
  {Barraza-Lopez}},\ and\ \bibinfo {author} {\bibfnamefont {J.}~\bibnamefont {Hu}},\ }\href {https://doi.org/https://doi.org/10.1002/qute.202100063} {\bibfield  {journal} {\bibinfo  {journal} {Adv. Quantum Technol.}\ }\textbf {\bibinfo {volume} {4}},\ \bibinfo {pages} {2100063} (\bibinfo {year} {2021})}\BibitemShut {NoStop}%
\bibitem [{\citenamefont {Rossiter}\ and\ \citenamefont {Bass}(2006)}]{Rossiter2006}%
  \BibitemOpen
  \bibfield  {author} {\bibinfo {author} {\bibfnamefont {P.~L.}\ \bibnamefont {Rossiter}}\ and\ \bibinfo {author} {\bibfnamefont {J.}~\bibnamefont {Bass}},\ }\href {https://onlinelibrary.wiley.com/doi/abs/10.1002/9783527603978.mst0033} {\bibfield  {journal} {\bibinfo  {journal} {Mater. Sci. Technol.}\ } (\bibinfo {year} {2006})}\BibitemShut {NoStop}%
\bibitem [{\citenamefont {Yuan}\ \emph {et~al.}(2024)\citenamefont {Yuan}, \citenamefont {Huang}, \citenamefont {Ma}, \citenamefont {Chen}, \citenamefont {Ren}, \citenamefont {Zhang}, \citenamefont {Feng}, \citenamefont {Zhu}, \citenamefont {Wang}, \citenamefont {He}, \citenamefont {Wu}, \citenamefont {Tan}, \citenamefont {Hao}, \citenamefont {Zhang}, \citenamefont {Liu}, \citenamefont {Liu}, \citenamefont {Liu}, \citenamefont {Cao}, \citenamefont {Chen},\ and\ \citenamefont {Lai}}]{Yuan2024}%
  \BibitemOpen
  \bibfield  {author} {\bibinfo {author} {\bibfnamefont {D.}~\bibnamefont {Yuan}}, \bibinfo {author} {\bibfnamefont {D.}~\bibnamefont {Huang}}, \bibinfo {author} {\bibfnamefont {X.}~\bibnamefont {Ma}}, \bibinfo {author} {\bibfnamefont {X.}~\bibnamefont {Chen}}, \bibinfo {author} {\bibfnamefont {H.}~\bibnamefont {Ren}}, \bibinfo {author} {\bibfnamefont {Y.}~\bibnamefont {Zhang}}, \bibinfo {author} {\bibfnamefont {W.}~\bibnamefont {Feng}}, \bibinfo {author} {\bibfnamefont {X.}~\bibnamefont {Zhu}}, \bibinfo {author} {\bibfnamefont {B.}~\bibnamefont {Wang}}, \bibinfo {author} {\bibfnamefont {X.}~\bibnamefont {He}}, \bibinfo {author} {\bibfnamefont {J.}~\bibnamefont {Wu}}, \bibinfo {author} {\bibfnamefont {S.}~\bibnamefont {Tan}}, \bibinfo {author} {\bibfnamefont {Q.}~\bibnamefont {Hao}}, \bibinfo {author} {\bibfnamefont {Q.}~\bibnamefont {Zhang}}, \bibinfo {author} {\bibfnamefont {Y.}~\bibnamefont {Liu}}, \bibinfo {author} {\bibfnamefont {Q.}~\bibnamefont {Liu}}, \bibinfo {author} {\bibfnamefont {Z.}~\bibnamefont
  {Liu}}, \bibinfo {author} {\bibfnamefont {C.}~\bibnamefont {Cao}}, \bibinfo {author} {\bibfnamefont {Q.}~\bibnamefont {Chen}},\ and\ \bibinfo {author} {\bibfnamefont {X.}~\bibnamefont {Lai}},\ }\href {https://doi.org/10.1103/PhysRevB.109.045113} {\bibfield  {journal} {\bibinfo  {journal} {Phys. Rev. B}\ }\textbf {\bibinfo {volume} {109}},\ \bibinfo {pages} {045113} (\bibinfo {year} {2024})}\BibitemShut {NoStop}%
\bibitem [{\citenamefont {Gao}\ \emph {et~al.}(2022)\citenamefont {Gao}, \citenamefont {Huang}, \citenamefont {Ren}, \citenamefont {Li}, \citenamefont {Wang}, \citenamefont {Yang}, \citenamefont {Li},\ and\ \citenamefont {Zhang}}]{Gao2022}%
  \BibitemOpen
  \bibfield  {author} {\bibinfo {author} {\bibfnamefont {F.}~\bibnamefont {Gao}}, \bibinfo {author} {\bibfnamefont {J.}~\bibnamefont {Huang}}, \bibinfo {author} {\bibfnamefont {W.}~\bibnamefont {Ren}}, \bibinfo {author} {\bibfnamefont {M.}~\bibnamefont {Li}}, \bibinfo {author} {\bibfnamefont {H.}~\bibnamefont {Wang}}, \bibinfo {author} {\bibfnamefont {T.}~\bibnamefont {Yang}}, \bibinfo {author} {\bibfnamefont {B.}~\bibnamefont {Li}},\ and\ \bibinfo {author} {\bibfnamefont {Z.}~\bibnamefont {Zhang}},\ }\href {https://doi.org/10.1103/PhysRevB.105.214434} {\bibfield  {journal} {\bibinfo  {journal} {Phys. Rev. B}\ }\textbf {\bibinfo {volume} {105}},\ \bibinfo {pages} {214434} (\bibinfo {year} {2022})}\BibitemShut {NoStop}%
\bibitem [{\citenamefont {Yue}\ \emph {et~al.}(2020)\citenamefont {Yue}, \citenamefont {Qian}, \citenamefont {Yang}, \citenamefont {Geng}, \citenamefont {Yi}, \citenamefont {Kumar}, \citenamefont {Shimada}, \citenamefont {Cheng}, \citenamefont {Chen}, \citenamefont {Wang}, \citenamefont {Weng}, \citenamefont {Shi}, \citenamefont {Wu},\ and\ \citenamefont {Feng}}]{Yue2020}%
  \BibitemOpen
  \bibfield  {author} {\bibinfo {author} {\bibfnamefont {S.}~\bibnamefont {Yue}}, \bibinfo {author} {\bibfnamefont {Y.}~\bibnamefont {Qian}}, \bibinfo {author} {\bibfnamefont {M.}~\bibnamefont {Yang}}, \bibinfo {author} {\bibfnamefont {D.}~\bibnamefont {Geng}}, \bibinfo {author} {\bibfnamefont {C.}~\bibnamefont {Yi}}, \bibinfo {author} {\bibfnamefont {S.}~\bibnamefont {Kumar}}, \bibinfo {author} {\bibfnamefont {K.}~\bibnamefont {Shimada}}, \bibinfo {author} {\bibfnamefont {P.}~\bibnamefont {Cheng}}, \bibinfo {author} {\bibfnamefont {L.}~\bibnamefont {Chen}}, \bibinfo {author} {\bibfnamefont {Z.}~\bibnamefont {Wang}}, \bibinfo {author} {\bibfnamefont {H.}~\bibnamefont {Weng}}, \bibinfo {author} {\bibfnamefont {Y.}~\bibnamefont {Shi}}, \bibinfo {author} {\bibfnamefont {K.}~\bibnamefont {Wu}},\ and\ \bibinfo {author} {\bibfnamefont {B.}~\bibnamefont {Feng}},\ }\href {https://doi.org/10.1103/PhysRevB.102.155109} {\bibfield  {journal} {\bibinfo  {journal} {Phys. Rev. B}\ }\textbf {\bibinfo {volume} {102}},\
  \bibinfo {pages} {155109} (\bibinfo {year} {2020})}\BibitemShut {NoStop}%
\bibitem [{\citenamefont {Singh}\ \emph {et~al.}(2024)\citenamefont {Singh}, \citenamefont {Kant~Mishra},\ and\ \citenamefont {Kumar~Ganguli}}]{Singh2024}%
  \BibitemOpen
  \bibfield  {author} {\bibinfo {author} {\bibfnamefont {H.}~\bibnamefont {Singh}}, \bibinfo {author} {\bibfnamefont {P.}~\bibnamefont {Kant~Mishra}},\ and\ \bibinfo {author} {\bibfnamefont {A.}~\bibnamefont {Kumar~Ganguli}},\ }\href {https://pubs.aip.org/aip/jap/article/136/7/073905/3308968} {\bibfield  {journal} {\bibinfo  {journal} {J. Appl. Phys.}\ }\textbf {\bibinfo {volume} {136}} (\bibinfo {year} {2024})}\BibitemShut {NoStop}%
\bibitem [{\citenamefont {Chen}\ \emph {et~al.}(2020)\citenamefont {Chen}, \citenamefont {Fei}, \citenamefont {Li}, \citenamefont {Wang}, \citenamefont {Luo}, \citenamefont {Yan}, \citenamefont {Lu}, \citenamefont {Tong}, \citenamefont {Song}, \citenamefont {Zhu}, \citenamefont {Zhang}, \citenamefont {Zhou}, \citenamefont {Zheng}, \citenamefont {Zhang}, \citenamefont {Lichtenstein}, \citenamefont {Katsnelson}, \citenamefont {Yin}, \citenamefont {Hao},\ and\ \citenamefont {Sun}}]{Chen2020}%
  \BibitemOpen
  \bibfield  {author} {\bibinfo {author} {\bibfnamefont {F.~C.}\ \bibnamefont {Chen}}, \bibinfo {author} {\bibfnamefont {Y.}~\bibnamefont {Fei}}, \bibinfo {author} {\bibfnamefont {S.~J.}\ \bibnamefont {Li}}, \bibinfo {author} {\bibfnamefont {Q.}~\bibnamefont {Wang}}, \bibinfo {author} {\bibfnamefont {X.}~\bibnamefont {Luo}}, \bibinfo {author} {\bibfnamefont {J.}~\bibnamefont {Yan}}, \bibinfo {author} {\bibfnamefont {W.~J.}\ \bibnamefont {Lu}}, \bibinfo {author} {\bibfnamefont {P.}~\bibnamefont {Tong}}, \bibinfo {author} {\bibfnamefont {W.~H.}\ \bibnamefont {Song}}, \bibinfo {author} {\bibfnamefont {X.~B.}\ \bibnamefont {Zhu}}, \bibinfo {author} {\bibfnamefont {L.}~\bibnamefont {Zhang}}, \bibinfo {author} {\bibfnamefont {H.~B.}\ \bibnamefont {Zhou}}, \bibinfo {author} {\bibfnamefont {F.~W.}\ \bibnamefont {Zheng}}, \bibinfo {author} {\bibfnamefont {P.}~\bibnamefont {Zhang}}, \bibinfo {author} {\bibfnamefont {A.~L.}\ \bibnamefont {Lichtenstein}}, \bibinfo {author} {\bibfnamefont {M.~I.}\ \bibnamefont
  {Katsnelson}}, \bibinfo {author} {\bibfnamefont {Y.}~\bibnamefont {Yin}}, \bibinfo {author} {\bibfnamefont {N.}~\bibnamefont {Hao}},\ and\ \bibinfo {author} {\bibfnamefont {Y.~P.}\ \bibnamefont {Sun}},\ }\href {https://doi.org/10.1103/PhysRevLett.124.236601} {\bibfield  {journal} {\bibinfo  {journal} {Phys. Rev. Lett.}\ }\textbf {\bibinfo {volume} {124}},\ \bibinfo {pages} {236601} (\bibinfo {year} {2020})}\BibitemShut {NoStop}%
\bibitem [{\citenamefont {Feng}\ \emph {et~al.}(2015)\citenamefont {Feng}, \citenamefont {Pang}, \citenamefont {Wu}, \citenamefont {Wang}, \citenamefont {Weng}, \citenamefont {Li}, \citenamefont {Dai}, \citenamefont {Fang}, \citenamefont {Shi},\ and\ \citenamefont {Lu}}]{Feng2017}%
  \BibitemOpen
  \bibfield  {author} {\bibinfo {author} {\bibfnamefont {J.}~\bibnamefont {Feng}}, \bibinfo {author} {\bibfnamefont {Y.}~\bibnamefont {Pang}}, \bibinfo {author} {\bibfnamefont {D.}~\bibnamefont {Wu}}, \bibinfo {author} {\bibfnamefont {Z.}~\bibnamefont {Wang}}, \bibinfo {author} {\bibfnamefont {H.}~\bibnamefont {Weng}}, \bibinfo {author} {\bibfnamefont {J.}~\bibnamefont {Li}}, \bibinfo {author} {\bibfnamefont {X.}~\bibnamefont {Dai}}, \bibinfo {author} {\bibfnamefont {Z.}~\bibnamefont {Fang}}, \bibinfo {author} {\bibfnamefont {Y.}~\bibnamefont {Shi}},\ and\ \bibinfo {author} {\bibfnamefont {L.}~\bibnamefont {Lu}},\ }\href {https://doi.org/10.1103/PhysRevB.92.081306} {\bibfield  {journal} {\bibinfo  {journal} {Phys. Rev. B}\ }\textbf {\bibinfo {volume} {92}},\ \bibinfo {pages} {081306} (\bibinfo {year} {2015})}\BibitemShut {NoStop}%
\bibitem [{\citenamefont {Xiong}\ \emph {et~al.}(2015)\citenamefont {Xiong}, \citenamefont {Kushwaha}, \citenamefont {Liang}, \citenamefont {Krizan}, \citenamefont {Hirschberger}, \citenamefont {Wang}, \citenamefont {Cava},\ and\ \citenamefont {Ong}}]{Xiong2015}%
  \BibitemOpen
  \bibfield  {author} {\bibinfo {author} {\bibfnamefont {J.}~\bibnamefont {Xiong}}, \bibinfo {author} {\bibfnamefont {S.~K.}\ \bibnamefont {Kushwaha}}, \bibinfo {author} {\bibfnamefont {T.}~\bibnamefont {Liang}}, \bibinfo {author} {\bibfnamefont {J.~W.}\ \bibnamefont {Krizan}}, \bibinfo {author} {\bibfnamefont {M.}~\bibnamefont {Hirschberger}}, \bibinfo {author} {\bibfnamefont {W.}~\bibnamefont {Wang}}, \bibinfo {author} {\bibfnamefont {R.~J.}\ \bibnamefont {Cava}},\ and\ \bibinfo {author} {\bibfnamefont {N.~P.}\ \bibnamefont {Ong}},\ }\href {https://doi.org/10.1126/science.aac6089} {\bibfield  {journal} {\bibinfo  {journal} {Science}\ }\textbf {\bibinfo {volume} {350}},\ \bibinfo {pages} {413} (\bibinfo {year} {2015})}\BibitemShut {NoStop}%
\bibitem [{\citenamefont {He}\ \emph {et~al.}(2014)\citenamefont {He}, \citenamefont {Hong}, \citenamefont {Dong}, \citenamefont {Pan}, \citenamefont {Zhang}, \citenamefont {Zhang},\ and\ \citenamefont {Li}}]{He2014}%
  \BibitemOpen
  \bibfield  {author} {\bibinfo {author} {\bibfnamefont {L.~P.}\ \bibnamefont {He}}, \bibinfo {author} {\bibfnamefont {X.~C.}\ \bibnamefont {Hong}}, \bibinfo {author} {\bibfnamefont {J.~K.}\ \bibnamefont {Dong}}, \bibinfo {author} {\bibfnamefont {J.}~\bibnamefont {Pan}}, \bibinfo {author} {\bibfnamefont {Z.}~\bibnamefont {Zhang}}, \bibinfo {author} {\bibfnamefont {J.}~\bibnamefont {Zhang}},\ and\ \bibinfo {author} {\bibfnamefont {S.~Y.}\ \bibnamefont {Li}},\ }\href {https://doi.org/10.1103/PhysRevLett.113.246402} {\bibfield  {journal} {\bibinfo  {journal} {Phys. Rev. Lett.}\ }\textbf {\bibinfo {volume} {113}},\ \bibinfo {pages} {246402} (\bibinfo {year} {2014})}\BibitemShut {NoStop}%
\bibitem [{\citenamefont {Hu}\ \emph {et~al.}(2016)\citenamefont {Hu}, \citenamefont {Tang}, \citenamefont {Liu}, \citenamefont {Liu}, \citenamefont {Zhu}, \citenamefont {Graf}, \citenamefont {Myhro}, \citenamefont {Tran}, \citenamefont {Lau}, \citenamefont {Wei},\ and\ \citenamefont {Mao}}]{Hu2016}%
  \BibitemOpen
  \bibfield  {author} {\bibinfo {author} {\bibfnamefont {J.}~\bibnamefont {Hu}}, \bibinfo {author} {\bibfnamefont {Z.}~\bibnamefont {Tang}}, \bibinfo {author} {\bibfnamefont {J.}~\bibnamefont {Liu}}, \bibinfo {author} {\bibfnamefont {X.}~\bibnamefont {Liu}}, \bibinfo {author} {\bibfnamefont {Y.}~\bibnamefont {Zhu}}, \bibinfo {author} {\bibfnamefont {D.}~\bibnamefont {Graf}}, \bibinfo {author} {\bibfnamefont {K.}~\bibnamefont {Myhro}}, \bibinfo {author} {\bibfnamefont {S.}~\bibnamefont {Tran}}, \bibinfo {author} {\bibfnamefont {C.~N.}\ \bibnamefont {Lau}}, \bibinfo {author} {\bibfnamefont {J.}~\bibnamefont {Wei}},\ and\ \bibinfo {author} {\bibfnamefont {Z.}~\bibnamefont {Mao}},\ }\href {https://doi.org/10.1103/PhysRevLett.117.016602} {\bibfield  {journal} {\bibinfo  {journal} {Phys. Rev. Lett.}\ }\textbf {\bibinfo {volume} {117}},\ \bibinfo {pages} {016602} (\bibinfo {year} {2016})}\BibitemShut {NoStop}%
\bibitem [{\citenamefont {Sankar}\ \emph {et~al.}(2017)\citenamefont {Sankar}, \citenamefont {Peramaiyan}, \citenamefont {Muthuselvam}, \citenamefont {Butler}, \citenamefont {Dimitri}, \citenamefont {Neupane}, \citenamefont {Rao}, \citenamefont {Lin},\ and\ \citenamefont {Chou}}]{Sankar2017}%
  \BibitemOpen
  \bibfield  {author} {\bibinfo {author} {\bibfnamefont {R.}~\bibnamefont {Sankar}}, \bibinfo {author} {\bibfnamefont {G.}~\bibnamefont {Peramaiyan}}, \bibinfo {author} {\bibfnamefont {I.~P.}\ \bibnamefont {Muthuselvam}}, \bibinfo {author} {\bibfnamefont {C.~J.}\ \bibnamefont {Butler}}, \bibinfo {author} {\bibfnamefont {K.}~\bibnamefont {Dimitri}}, \bibinfo {author} {\bibfnamefont {M.}~\bibnamefont {Neupane}}, \bibinfo {author} {\bibfnamefont {G.~N.}\ \bibnamefont {Rao}}, \bibinfo {author} {\bibfnamefont {M.-T.}\ \bibnamefont {Lin}},\ and\ \bibinfo {author} {\bibfnamefont {F.~C.}\ \bibnamefont {Chou}},\ }\href {https://doi.org/10.1038/srep40603} {\bibfield  {journal} {\bibinfo  {journal} {Sci. Rep.}\ }\textbf {\bibinfo {volume} {7}},\ \bibinfo {pages} {40603} (\bibinfo {year} {2017})}\BibitemShut {NoStop}%
\bibitem [{\citenamefont {Yang}\ \emph {et~al.}(2020)\citenamefont {Yang}, \citenamefont {Qian}, \citenamefont {Yan}, \citenamefont {Li}, \citenamefont {Song}, \citenamefont {Wang}, \citenamefont {Yi}, \citenamefont {Feng}, \citenamefont {Weng},\ and\ \citenamefont {Shi}}]{Yang2020}%
  \BibitemOpen
  \bibfield  {author} {\bibinfo {author} {\bibfnamefont {M.}~\bibnamefont {Yang}}, \bibinfo {author} {\bibfnamefont {Y.}~\bibnamefont {Qian}}, \bibinfo {author} {\bibfnamefont {D.}~\bibnamefont {Yan}}, \bibinfo {author} {\bibfnamefont {Y.}~\bibnamefont {Li}}, \bibinfo {author} {\bibfnamefont {Y.}~\bibnamefont {Song}}, \bibinfo {author} {\bibfnamefont {Z.}~\bibnamefont {Wang}}, \bibinfo {author} {\bibfnamefont {C.}~\bibnamefont {Yi}}, \bibinfo {author} {\bibfnamefont {H.~L.}\ \bibnamefont {Feng}}, \bibinfo {author} {\bibfnamefont {H.}~\bibnamefont {Weng}},\ and\ \bibinfo {author} {\bibfnamefont {Y.}~\bibnamefont {Shi}},\ }\href {https://doi.org/10.1103/PhysRevMaterials.4.094203} {\bibfield  {journal} {\bibinfo  {journal} {Phys. Rev. Mater.}\ }\textbf {\bibinfo {volume} {4}},\ \bibinfo {pages} {094203} (\bibinfo {year} {2020})}\BibitemShut {NoStop}%
\end{thebibliography}
%

\end{document}